\documentclass[onecolumn,showpacs,preprintnumbers,amsmath,amssymb,prd,
floatfix,showkeys,11pt]{revtex4}

\usepackage{graphics}
\usepackage{graphicx}
\usepackage{epsfig}
\usepackage{color}
\usepackage{longtable}
\usepackage{hyperref}
\usepackage{latexsym}
\usepackage{amsmath}
\usepackage{amsthm}
\usepackage{amsfonts}
\usepackage{amssymb}
\usepackage{dsfont}
\usepackage{bm}
\newcommand{\ov}[1]{\overline{#1}}
\newcommand{\be}{\begin{equation}}
\newcommand{\ee}{\end{equation}}
\newcommand{\bey}{\begin{eqnarray}}
\newcommand{\eey}{\end{eqnarray}}
\newcommand{\tr}{\text{Tr}}
\newcommand{\ot}{\ensuremath{\frac{1}{2}}}
\newcommand{\DCI}{\ensuremath{D_\srm{CI}} }
\newcommand{\unitmatrix}{\mathds{1}}
\newcommand{\VEC}[1]{\ensuremath{\textrm{\boldmath{$#1$}}}}
\newcommand{\la}{\langle}
\newcommand{\ra}{\rangle}

\newcommand{\g}{\gamma}
\newcommand{\wm}{\phantom{-}}
\newcommand{\srm}[1]{\textrm{\scriptsize{#1}}}
\newcommand{\MSbar}{\ensuremath{\overline{\textrm{MS}}}}

\begin{document}
%
\title{Hadron Spectroscopy with Dynamical Chirally Improved Fermions}
\author{Christof Gattringer,$^1$
Christian Hagen,$^2$
C. B.~Lang,$^1$ 
Markus Limmer,$^1$ 
Daniel Mohler,$^1$ and
Andreas Sch\"afer$^2$\\
\vspace*{2mm}
(BGR [Bern-Graz-Regensburg] Collaboration)
\vspace*{2mm}\\}

\affiliation{
$^1$Institut f\"ur Physik, FB Theoretische Physik, Universit\"at
Graz, A--8010 Graz, Austria\\
$^2$Institut f\"ur Theoretische Physik, Universit\"at
Regensburg, D--93040 Regensburg, Germany
\vspace*{3mm}
}

\pacs{11.15.Ha, 12.38.Gc}
\keywords{Hadron spectroscopy, dynamical fermions}

\date{\today}

\begin{abstract}

We simulate two dynamical, mass degenerate light quarks on $16^3\times32$
lattices with a spatial extent of $2.4$ fm using the Chirally
Improved Dirac operator.  The simulation method, the implementation of the action and
signals of equilibration are  discussed in detail. Based on the
eigenvalues of the Dirac operator we discuss some qualitative features of our
approach.  Results for ground state masses of pseudoscalar and vector mesons
as well as for the nucleon and delta baryons are presented.

\end{abstract}

\maketitle


\section{Introduction\label{sec:intro}}

Lattice Dirac operators that obey the so-called Ginsparg-Wilson (GW) relation
\cite{GiWi82} implement a lattice version of the chiral symmetry transformations
\cite{Lu98}. Presently only one explicit formulation of lattice fermions, the
overlap  Dirac operator \cite{Ne98a,Ne98b}, is GW exact in that sense. There
are, however, several formulations approaching GW exactness in various ways.
Among them is the domain-wall formulation \cite{Ka92,FuSh95},  which approaches
the overlap operator in the limit of infinite extent of an artificial 5-th
dimension. Another one is the so-called perfect Dirac operator
\cite{HaNi94,HaHaNi05}, which if constructed explicitly, would obey the GW
condition, and which has been approximated by a parameterized fixed-point form.
Here we discuss a simulation with the so-called Chirally Improved (CI) Dirac
operator \cite{Ga01a,GaHiLa00}, which is also a parameterization of a Dirac
operator obeying the GW relation approximately.

Advantages of GW exact fermions are that there is no additive mass
renormalization and thus no spurious zero modes at non-zero quark masses.
Operators are protected by chiral symmetry which is convenient for the
determination of certain matrix elements. Technically the GW exact overlap
operator involves taking the square root of a simpler kernel operator (e.g., the
Wilson operator), which is computationally roughly two orders of magnitude more
expensive than simulations with non-GW-operators. This is not only due to the
technical implementation (through, e.g., polynomial series or rational
functions) but also due to tunneling problems between sectors of different
topology. Therefore, only a few groups have attempted to implement dynamical
overlap fermions
\cite{KaAoFu06,MaFuHa06,AoFuHa08,NoAoFu07,CuKrLi05,CuKrAr05,CuKrLi07,
CuKrLi08,DeSc05a,DeSc05b,FoKaSz05,EgFoKa05}.

On the other hand, GW-type operators, fulfilling the GW condition in some
approximation, although more expensive than simple Wilson-Dirac operators, have
been studied in quenched calculations within the BGR collaboration for some
time. There we demonstrated that at least for baryon masses the
$\mathcal{O}(a^2)$ corrections are quite small \cite{GaGoHa03a} and that field
renormalization constants behave almost like in the chirally symmetric case
\cite{GaHuLa05a}. Motivated by these results we have started to implement CI
fermions for a dynamical simulation on smaller lattices \cite{LaMaOr05c} and are
now presenting details and results of our simulations on larger lattices.

First results involving dynamical CI fermions on $16^3\times 32$ lattices were
published in \cite{FrGaLa07, La08}, and results for smaller lattices can be
found in \cite{LaMaOr05c, LaMaOr05b}. In this paper we concentrate on  technical
aspects of the simulation and present first results for the hadron mass spectrum
for three sets of parameters, corresponding to three different pion masses,
giving an overview of the current project status. We start with an
explanation of all the technicalities, i.e., simulation details and
equilibration behavior, followed by the first analysis results for ground state
masses of mesons and baryons. We finish with a discussion of the results and a
summary.


\section{Setup \label{sec:setup} and simulation}

\subsection{CI Dirac operator and action}

For the fermions we use the so-called Chirally Improved (CI) Dirac operator
$\DCI$ \cite{Ga01a,GaHiLa00}. It obeys chiral symmetry only approximately,
depending on the truncation in the extent of the interaction terms. Plugging a
general  ansatz into the Ginsparg-Wilson equation leads to a set of algebraic
equations for the coefficients, which can be solved to obtain $\DCI$. The paths
and coefficients used in our simulation are given in Appendix \ref{app:DCI}.
Whereas in the quenched simulations the \DCI coefficients were adapted to the
values of the gauge coupling such as to have (almost) no mass renormalization,
we now decided to use the same \DCI parameters for all dynamical runs. This
implies an additive mass renormalization, i.e., the ``mass parameter'' $m_0$ does
not give the bare mass directly. We adjust the value  of $m_0$ such as to get
suitable PCAC masses (also called AWI-masses since their definition comes from
the axial Ward identity). The numbers will be discussed in more detail below.

It was shown in a quenched calculation using $\DCI$ \cite{GaHiLa00}, that the
L{\"u}scher-Weisz gauge action \cite{LuWe85} produces smoother gauge
configurations than the Wilson gauge action, and thus is used in our simulation.
For completeness we also list details of the gauge action in  Appendix \ref{app:LW}.

Another important ingredient in our simulation is smearing, since the smearing
procedure results in better chiral properties of the operator, as can be seen
from the eigenvalue spectrum of the Dirac operator  \cite{DuHo04}. In earlier
quenched studies with $\DCI$ so-called HYP smearing \cite{HaKn01} was applied.
Since such a smearing procedure is not differentiable and thus not well suited
for Hybrid Monte-Carlo simulations, we decided to use the ``differentiable''
stout-smearing \cite{MoPe04}. In our simulation we have been using one level of
stout-smearing, such that the value of the plaquette is maximized. The stout
smearing is considered to be part of the definition of the full Dirac operator.

More recently, other suggestions for efficient differentiable smearing methods
have been published \cite{HaHoSc07,Du07}. For consistency we continued to use
the stout type smearing with which we started our study.

\subsection{Run parameters}

For the simulation presented here we use lattices of size $16^3\times 32$ at
three different values of the gauge coupling $\beta_1$ and the bare mass
parameter $m_0$, all of which can be found in Tab.~\ref{tab:parameters}. The
physical volume is always $\sim 2.4$ fm. The pion mass ranges from approximately
$530$ MeV down to $320$ MeV. The total number of  gauge configurations produced,
$N_\srm{conf}$, can also be found in Tab.~\ref{tab:parameters}.

\begin{table}[tb]\begin{center}
\begin{tabular}{cccccccc} \hline\hline
~Run~ & $m_0$ & $\beta_1$ & $\beta_2$ & $\beta_3$ & $m_\srm{HB}$ & ~$N_\srm{conf}$~ &
$P_\srm{acc}$\\
\hline
A & ~$-0.050$~ &  $4.70$  & $-0.3941$   & $-0.06063$   &  $\sqrt{0.03}$  & $100$ & $0.904$
\\
B & ~$-0.060$~ &  $4.65$  & $-0.3899$   & $-0.05998$   &  $\sqrt{0.02}$  & $200$ & $0.911$
\\
C & ~$-0.077$~ & ~$4.58$~ & ~$-0.3841$~ & ~$-0.05908$~ & ~$\sqrt{0.02}$~ & $200$ &
~$0.858$~ \\
\hline\hline
\end{tabular}
\caption{The parameters for the different runs. The number of pseudofermions
$N_\srm{PF}=2$, the total length of the trajectories is $1$ in HMC time units.
In the 6-th column the parameter for the Hasenbusch mass preconditioning is
given (see Sect.\ \ref{subsec:algorithm} for more details).
\label{tab:parameters}}
\end{center}\end{table}

\subsection{Algorithm\label{subsec:algorithm}}

The algorithm we use for generating our gauge configurations is a Hybrid
Monte-Carlo (HMC) \cite{DuKePe87} algorithm plus some additional
features.  HMC seems to be the most suitable algorithm for our goal.

For the HMC we need a generalization of the Hamiltonian evolution for a
system of classical mechanics in a ficticious HMC time to our system of fields $U_{n,\mu}$. For
that purpose we introduce traceless hermitian matrices $P_{n,\mu} \in
\texttt{su}(3)$ which act as conjugate momenta for the $U_{n,\mu}$, with
$n,\mu$ being the lattice site and the direction of the link,
respectively. We now can define the time derivative of $U_{n,\mu}$ as
\be
\dot{U}_{n,\mu} = i\, P_{n,\mu}\, U_{n,\mu}\ .
\ee
Then, a Hamiltonian $H$ can be defined as
\be
H = \frac{1}{2} \sum_{n,\mu} \tr\left(P_{n,\mu}^2\right) + S_\srm{g} + \phi^\dagger
(D^\dagger
D)^{-1} \phi\ ,
\ee
where $S_\srm{g}$ denotes the gauge action and 
$\phi$ is the pseudofermion field. The equation of motion for $P$ is obtained via
the relation $\dot{H}=0$,
\be
\dot{H} = \sum_{n,\mu} \tr\left(P_{n,\mu} \dot{P}_{n,\mu}\right) + \dot{S_\srm{g}} +
\phi^\dagger
\frac{d}{dt} (D^\dagger D)^{-1} \phi = 0\ ,
\ee
which gives the evolution equation in HMC time $\dot{P}=f(U,\dot{U},P)$.
Evaluating this function for, e.g., Wilson or staggered quarks is not
complicated since such types of quarks involve only one link field $U_{n,\mu}$
connecting neighboring sites. In our case, however, paths up to length four,
coming from $\DCI$, have to be considered. A more detailed description of the
procedure can be found in \cite{Or06}. For the evolution in HMC time we used the
reversible and area preserving leapfrog integration scheme.

To be able to go to smaller quark masses we utilize Hasenbusch mass
preconditioning \cite{Ha01a}. The basic idea is to split the pseudofermion
action into two (or more) parts, separating the small and the large eigenvalues
of the Dirac matrix. In our case we always use two pseudofermions. The parameter
$m_\srm{HB}$, which amounts to an additional mass, is deduced from an educated
guess \cite{Or06}.  Using $N_\srm{PF}$ pseudofermions, the mass shift is given
by
\be
m_\srm{HB}^{(i)} = \left\{
\begin{array}{cl}
\left({2^{N_\srm{PF}-i}\: \lambda_\srm{min}^i }\right)^{1/N_\srm{PF}} & \ ,
\ 1\le i <  N_\srm{PF} \\
&\\
0 & \ ,\ i=N_\srm{PF}
\end{array}\ . \right.
\ee
Here, $\lambda_\srm{min}$ is the assumed smallest eigenvalue of the Dirac
matrix.

\begin{figure}[tb]\begin{center}
\includegraphics[width=8cm,clip]{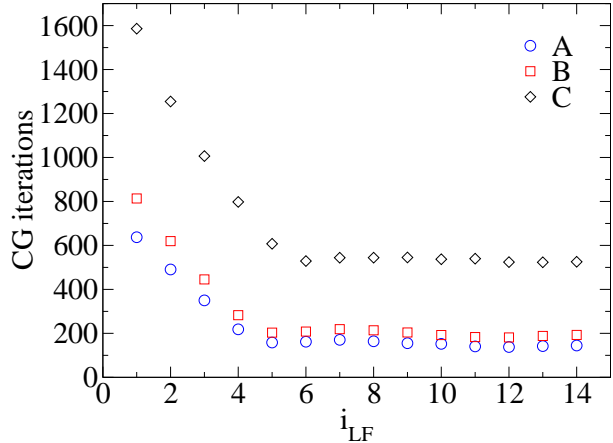}
\caption{The average number of needed conjugate gradient
iterations is plotted against the first leapfrog steps for each run.}
\label{fig:cg} 
\end{center}\end{figure}

For the inversion of $D^\dagger D$ we use the standard conjugate gradient (CG)
inverter. These inversions take by far most of the computer time. Thus, several
attempts were made to increase the performance of this part of our code. First
of all we use a chronological inverter by minimal residue extrapolation
\cite{BrIvLe97}, taking into account $12$ previous solutions. In
Fig.~\ref{fig:cg} we plot the number of conjugate gradient iterations against
the leapfrog step $i_\srm{LF}$. What we see is a rapid decrease in the CG
iteration number when more previous solutions become available. However, we find
that a plateau is reached already at $i_\srm{LF}=5$. The overhead caused by the $8$
additional matrix vector multiplications is negligible, however.

In a recent paper \cite{DuFoHo08} D{\"u}rr et al.~presented a mixed precision
inverter for the Dirac matrix. In order to ensure reversibility in the molecular
dynamics (MD) evolution one should work  with double precision accuracy. The
method suggested there allows to iteratively improve the inversion accuracy,
working partly with single precision and thus faster arithmetic. We choose a
final accuracy of $\varepsilon = 10^{-7}$. The gain in run-time per gauge
configuration was, e.g., about $33$\% for run C.

\subsection{Autocorrelation time}

\begin{figure}[tb]
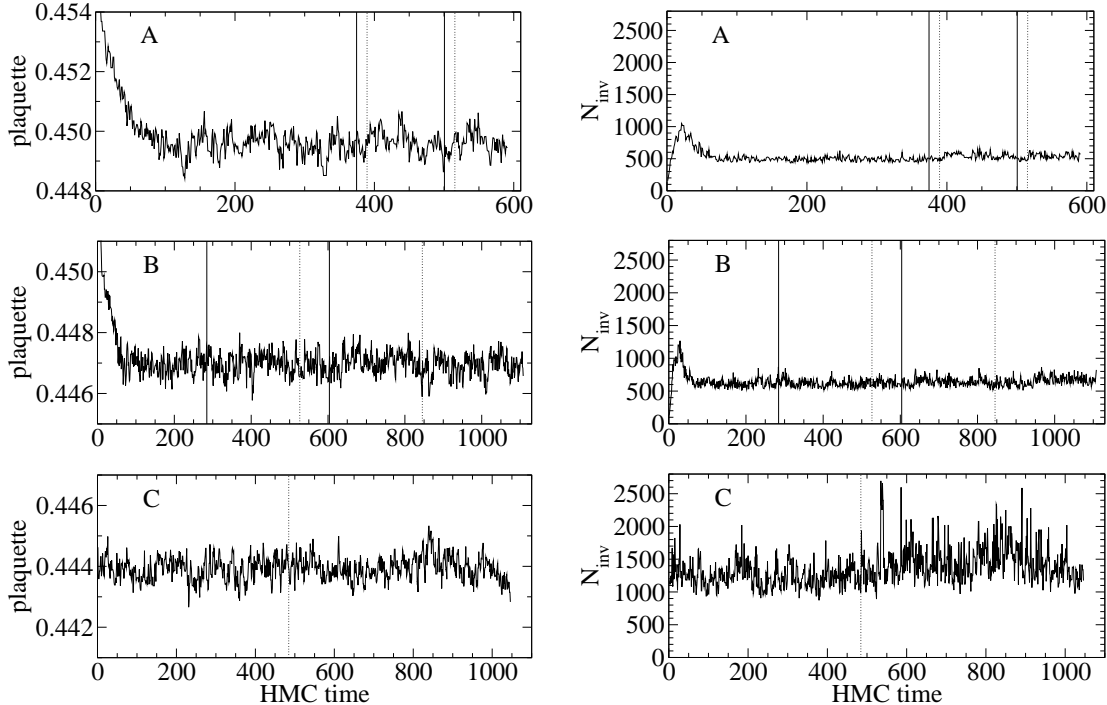
\begin{center}
\includegraphics[width=7cm,clip]{./figs/plaq_A.eps}\hspace*{5mm}
\includegraphics[width=7cm,clip]{./figs/final_cg_A.eps}\\[3mm]
\includegraphics[width=7cm,clip]{./figs/plaq_B.eps}\hspace*{5mm}
\includegraphics[width=7cm,clip]{./figs/final_cg_B.eps}\\[3mm]
\includegraphics[width=7cm,clip]{./figs/plaq_C.eps}\hspace*{5mm}
\includegraphics[width=7cm,clip]{./figs/final_cg_C.eps}\\
\caption{L.h.s.: The spatially averaged plaquette against the HMC time, from top
to bottom we plot runs A,  B and C. The dashed line in the plots indicate a
change in the algorithm: From that point on we changed to the mixed precision
inverter and used Hasenbusch mass preconditioning. In run C the Hasenbusch mass
preconditioning was used from the beginning, we only changed to the mixed
precision inverter. The full lines in run A and run B indicate a split in the
particular run into two separate trajectories to increase the production of
gauge configurations per (real) time. R.h.s.: The number of CG iterations,
$N_\srm{inv}$ in the accept/reject step, notation like l.h.s. }\label{fig:plaq}
\end{center}\end{figure}

A measure for the statistical efficiency of an observable $O$ is the integrated
autocorrelation time $\tau_\srm{int}$, defined by
\be \label{eq:autocorr1}
\tau_\srm{int} = \frac{1}{2} + \sum_{t=1}^\infty \frac{\Gamma(t)}{\Gamma(0)}\ ,
\ee
where the autocorrelation function $\Gamma$ is given by
\be
\Gamma(t) = \Big\langle \big(O(t_0) - \langle O \rangle\big) \big(O(t_0+t) - \langle O
\rangle\big) \Big\rangle\ .
\ee
In practice, the sum (\ref{eq:autocorr1}) has to be truncated at some upper
value $t_\srm{max}$, which we choose at that point where the autocorrelation
data becomes noisy. We discuss several observables to be able to figure out the
point of equilibration and the statistical independence of our measurements.

On the l.h.s.~of Fig.~\ref{fig:plaq} we plot the plaquette values for the three
runs. One can clearly see that the runs A and B, starting from ``cool'' quenched
configurations, are equilibrated after roughly $\mathcal{O}(100)$
configurations. Run C does not show a significant equilibration process for the
following reason. We started from a configuration B and slowly changed the
parameters $\beta_1$ and $m_0$ to the values of run C. This was the starting
configuration of the new run sequence C. 

Another indicator of equilibrium behavior is the number of CG-steps in the final
accept/reject step of the MD evolution, $N_\srm{inv}$. We show these numbers on 
the r.h.s.~of Fig.~\ref{fig:plaq}. Here one can also see that the runs are
equilibrated after the above mentioned number of HMC updates. Based on these
observations, in our analysis we discarded the first 100, 115, and 50
configurations for runs A, B, and C, respectively.

\begin{figure}[tb]
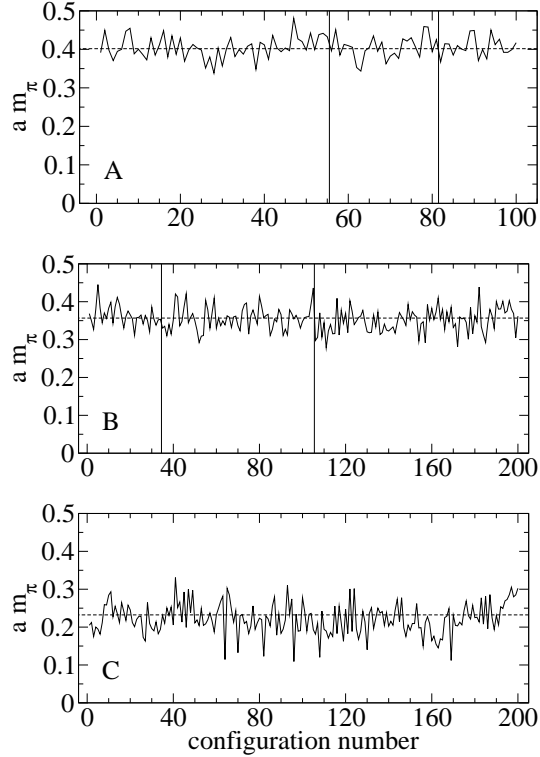
\begin{center}
\includegraphics[width=7cm,clip]{./figs/single_conf_masses_A.eps}\\[3mm]
\includegraphics[width=7cm,clip]{./figs/single_conf_masses_B.eps}\\[3mm]
\includegraphics[width=7cm,clip]{./figs/single_conf_masses_C.eps}\\
\caption{Time histories of the pion mass determined from single configurations
using the interpolator $\ov{u}_w\gamma_5 d_w$ (see 
Sect.\ \ref{sec_hadron_interpolators}). The horizontal dashed line indicates the mass
value obtained from fits to  the $ww$ propagators in the range $t=4-15$ (run A,
B) or $t=5-15$ (run C). The vertical lines in runs A and B indicate a split into
separate sequences such as to enhance statistics by parallel runs.}
\label{fig:singlepion}
\end{center}\end{figure}

Starting thus with the equilibrated configuration, we computed the  integrated
autocorrelation time $\tau_\srm{int}$ for the plaquette values and for
$N_\srm{inv}$. The resulting numbers are given in Tab.~\ref{tab:autocorr} and
are all below $5$. Therefrom we decided to analyze every 5-th configuration,
i.e., configurations separated by $5$ units of HMC time.

In Fig.~\ref{fig:singlepion} we show the history of the pion mass,  calculated
separately for each analyzed configuration. Details of the used pion
interpolator are discussed in the spectroscopy section. No noticeable
correlation can be found in the plots.

\begin{table}[tb]\begin{center}
\begin{tabular}{cccc} \hline\hline
Run~ & ~$N_\srm{equi}$~ & ~$\tau_\srm{int}(\text{plaq.})$~ &
~$\tau_\srm{int}(N_\srm{inv})$  \\ \hline
A & $100$ & $3.5$ & $4.2$  \\
B & $115$ & $2.4$ & $2.7$  \\
C & ~$50$  & $3.7$ & $3.6$  \\
\hline\hline
\end{tabular}
\caption{Integrated autocorrelation times for the three runs. $N_\srm{equi}$ is
the number of configurations skipped after the start.
\label{tab:autocorr}}
\end{center}\end{table}

\subsection{The change in the Hamiltonian}

\begin{figure}[tb]
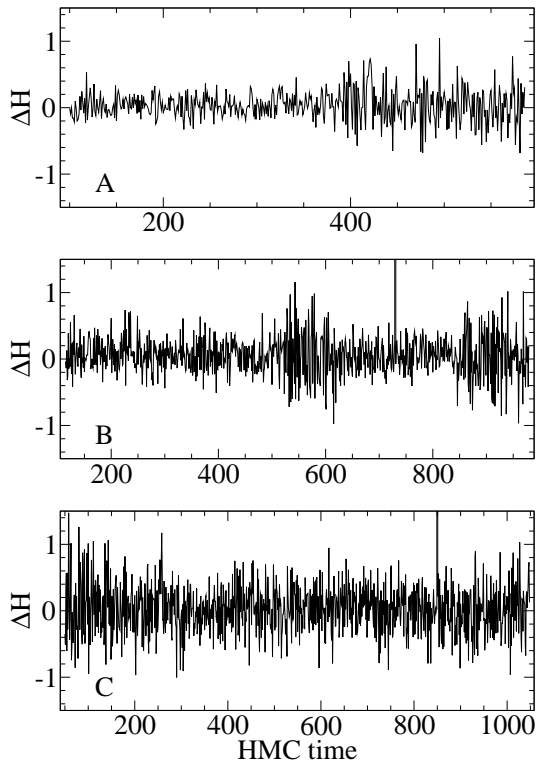
\begin{center}
\includegraphics[width=7cm,clip]{./figs/delta_s_A.eps}\\[3mm]
\includegraphics[width=7cm,clip]{./figs/delta_s_B.eps}\\[3mm]
\includegraphics[width=7cm,clip]{./figs/delta_s_C.eps}\\
\caption{We plot $\Delta H$ against the HMC time  starting from the point of equilibration
(runs A, B and C are ordered from top to bottom).}
\label{fig:delta_s}
\end{center}\end{figure}

Since we introduced the conjugate momenta $P$, we describe a microcanonical
ensemble of a classical system with a Hamiltonian $H$. For exact solutions of
the equations of motion  (MD equations) the Hamiltonian would be a constant of
motion and the configurations all would lie on a surface of constant energy.
Thus, each created configuration would be accepted. However, due to the
discretization with an MD time step $\delta_t$ numerical errors are introduced
and the Hamiltonian energy is not invariant. We denote the change as $\Delta H$.
Each calculated gauge configuration is then accepted with a probability
$e^{-\Delta H}$. The area preserving property of MD leads to an inequality
\cite{Cr88},
\be
e^{\langle -\Delta H\rangle} \le \left\langle e^{-\Delta H}\right\rangle = 1\ .
\ee
Due to this inequality $\langle\Delta H\rangle$ has to be positive and this is
indeed the case in our simulations (cf.~Tab.~\ref{tab:delta_s} and
Fig.~\ref{fig:delta_s} for our values). For run B we have a quite large value of
$\langle\Delta H\rangle$, coming from a huge spike in $\Delta H$ in
configuration $730$ which is of the order of $\mathcal{O}(10^5)$ bigger than the
rest. Also in run C we have a spike at configuration $850$, being about
$\mathcal{O}(10^3)$ bigger than the other values. Such spikes have already been
observed in other simulations with dynamical fermions \cite{NaAoFu04,AoFuHa08}.
Two possible reasons can cause such a spike. One is the instability of HMC for
large step sizes in the MD evolution, cf.~Ref.~\cite{JoPeKe00}. The other one,
and this is most likely the case here, is that the Dirac operator can develop
very small eigenvalues which lead to these spikes in the derivative of the
action.

\begin{figure}[tb]\begin{center}
\includegraphics[width=7cm,clip]{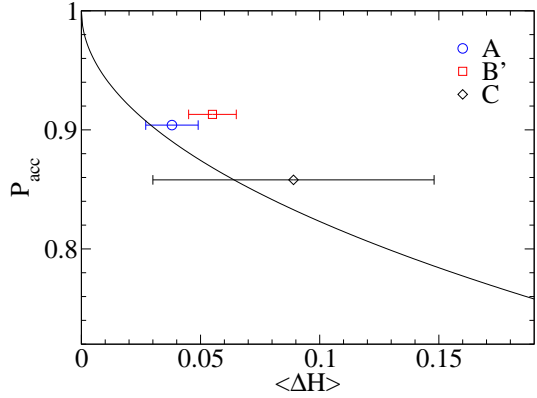}
\caption{$P_\srm{acc}$ vs.~$\langle \Delta H \rangle$. The black line corresponds to
$\text{erfc}(\sqrt{\Delta H}/2)$.
\label{fig:erfc}}
\end{center}\end{figure}

We want to conclude with a remark on the relation between $\Delta H$ and the
acceptance rate. In Fig.~\ref{fig:erfc} we plot the acceptance rate against the
averaged $\Delta H$. In our case, at least run A and run C are lying (within
error bars) on the predicted curve \cite{GuIrKa90},
\be
P_\srm{acc} = \text{erfc}\left( \frac{\sqrt{\Delta H}}{2} \right)\ ,
\ee
where erfc is the complementary error function.

\begin{table}[tb]\begin{center}
\begin{tabular}{lccc} \hline\hline
Run  & $\langle \Delta H \rangle$ &
~$e^{-\langle\Delta H \rangle}$~ & $\langle e^{-\Delta H} \rangle$   \\
\hline
A   & $0.038(11) $  & $0.963$ & $0.989(11)$ \\
B   & $2.01(1.95)$  & $0.134$ & $0.986(10)$\\
B'  & $0.055(10) $  & $0.947$ & $0.988(10)$ \\
C   & $0.089(59) $  & $0.915$ & $1.034(12)$\\
\hline\hline
\end{tabular}
\caption{Averages of $\Delta H$ and their exponentials for each run. We only
included the equilibrated configurations in our calculations. The 3-rd row
contains the data of run B without including configuration $730$, which is
responsible for the spike in $\Delta H$. 
\label{tab:delta_s}}
\end{center}\end{table}

\subsection{Dirac eigenvalues}

\begin{figure}[tp]
\begin{center}
\includegraphics[width=6.5cm,clip]{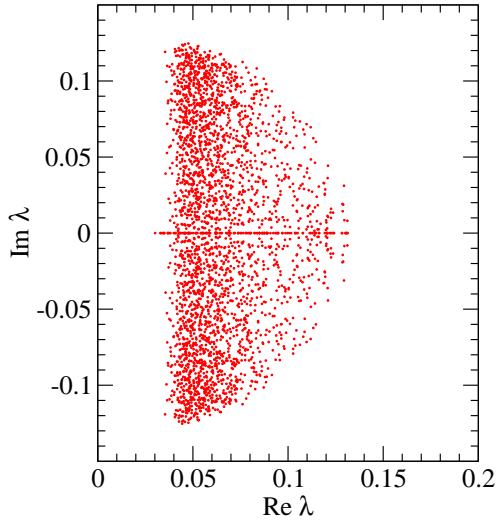}
\caption{The smallest 150 eigenvalues superimposed for 20\% of the configurations of
run A.}
\label{fig:eigenvalues_1}
\end{center}\end{figure}

An indicator of the ``GW quality'' of the Dirac operator is its eigenvalue
distribution in the complex plane. Whereas Dirac operators obeying the GW
condition in its simplest form have eigenvalues on a unit circle centered at 1,
approximate GW operators like \DCI deviate from that simple shape showing some
scattering of the eigenvalues. Figure \ref{fig:eigenvalues_1} shows the (in
absolute value) smallest 150 eigenvalues superimposed for 20\% of the
configurations of run A. Obviously the fluctuation is predominantly towards
values inside the unit circle and so-called exceptional configurations
(exceptionally small eigenvalues) are suppressed.

\begin{figure}[tp]
\begin{center}
\includegraphics*[width=7cm,clip]{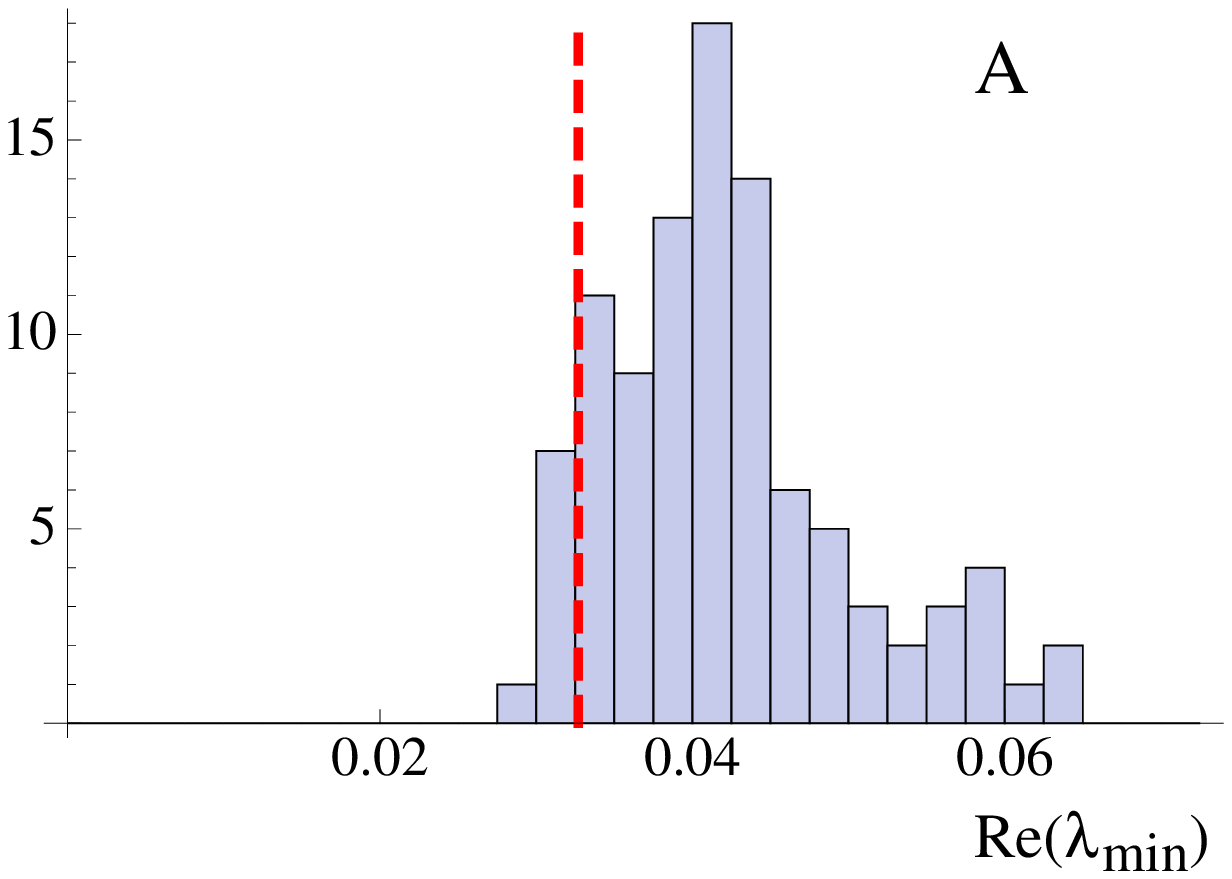}\hspace*{5mm}
\includegraphics*[width=7cm,clip]{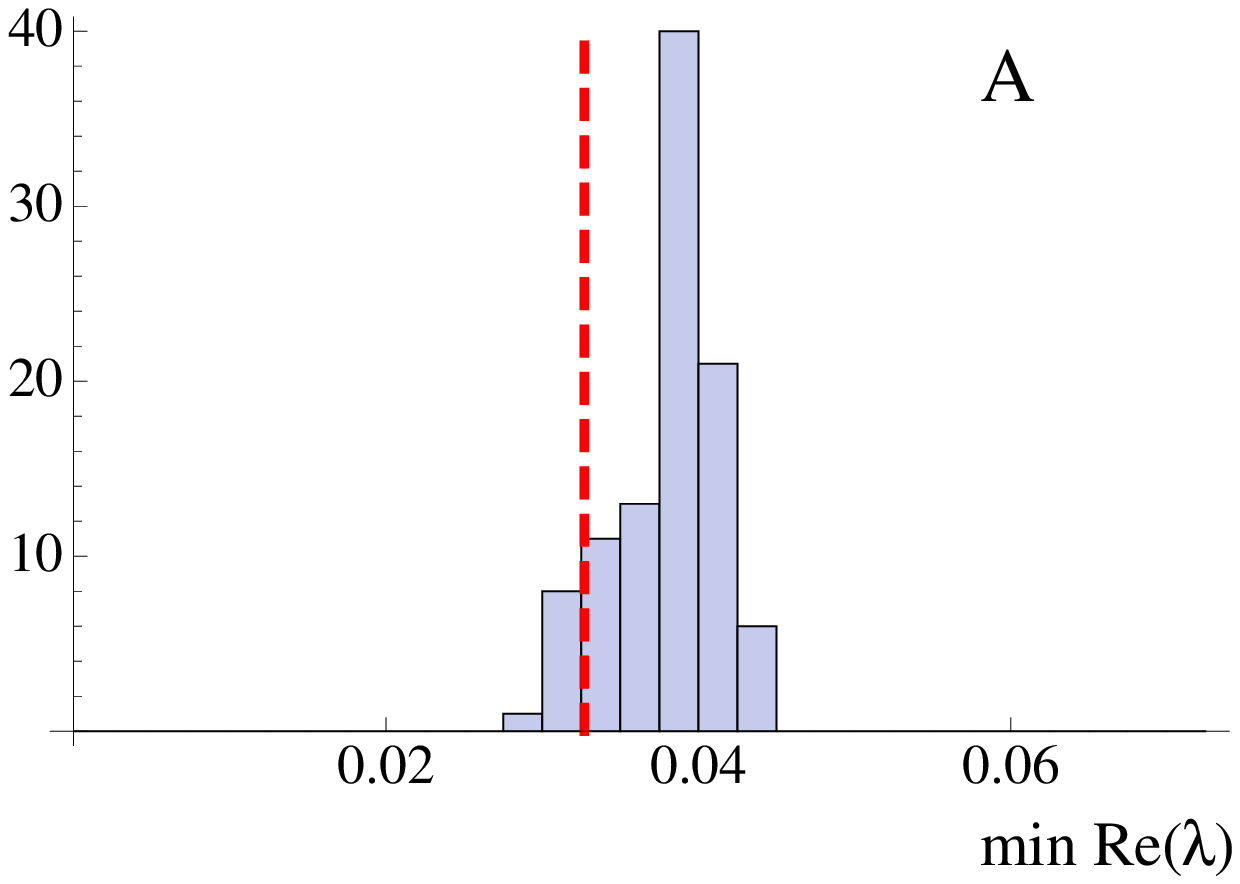}\\
\includegraphics*[width=7cm,clip]{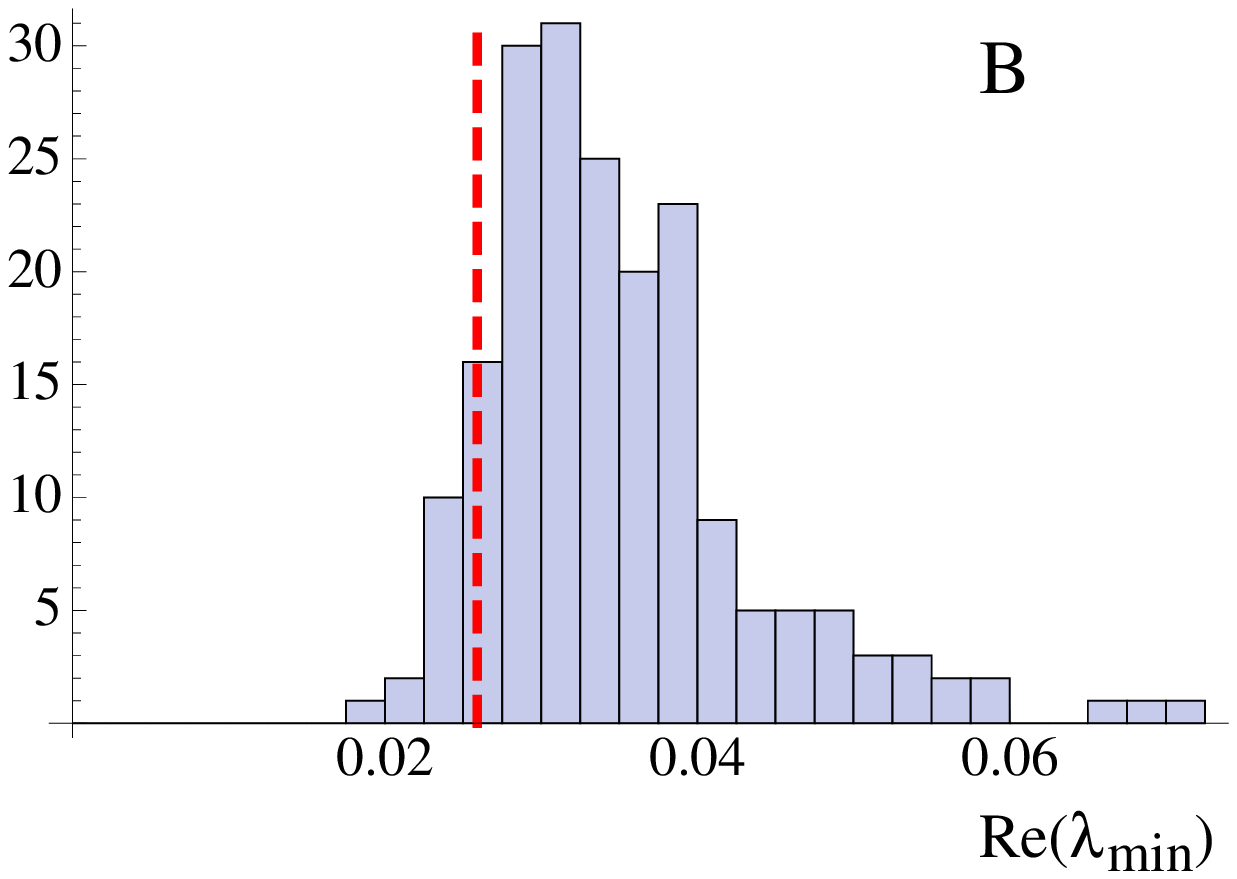}\hspace*{5mm}
\includegraphics*[width=7cm,clip]{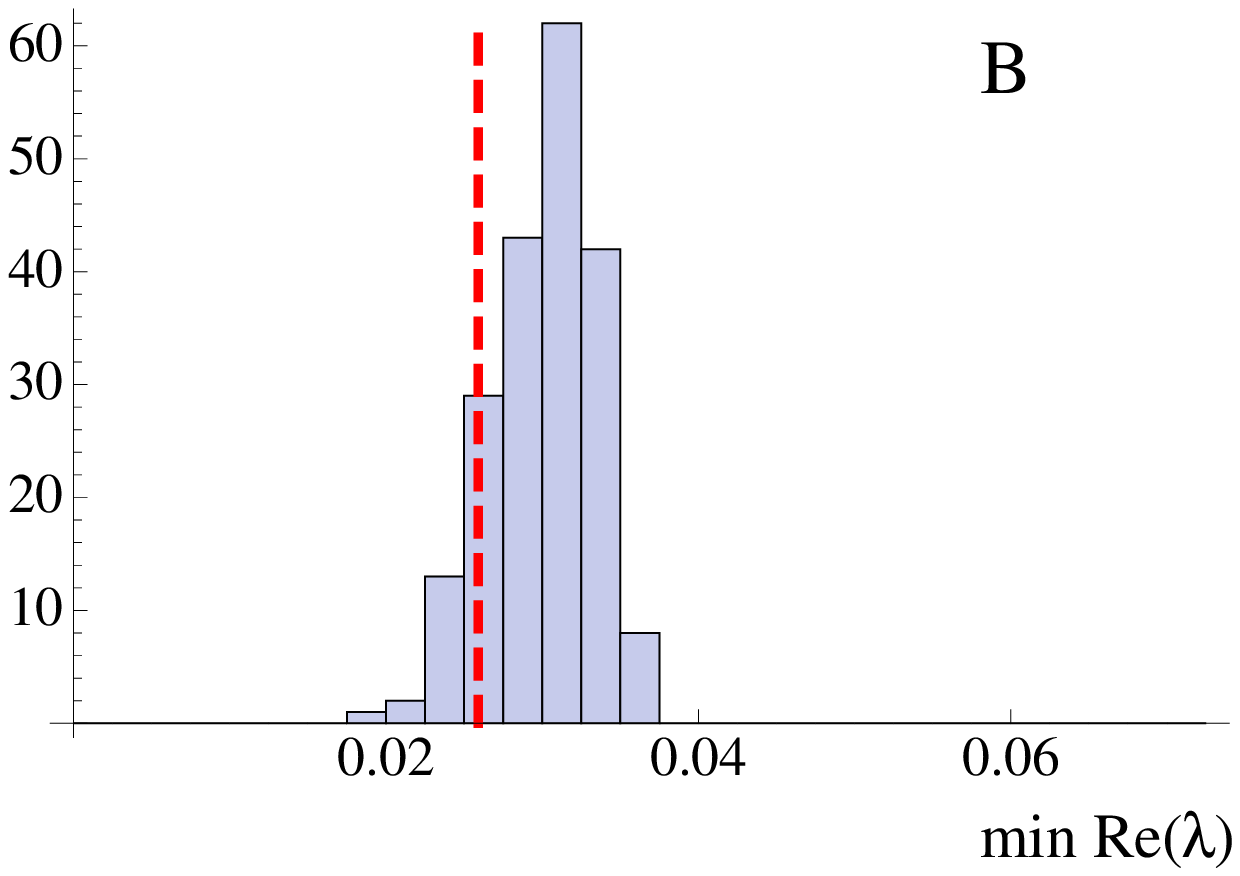}\\
\includegraphics*[width=7cm,clip]{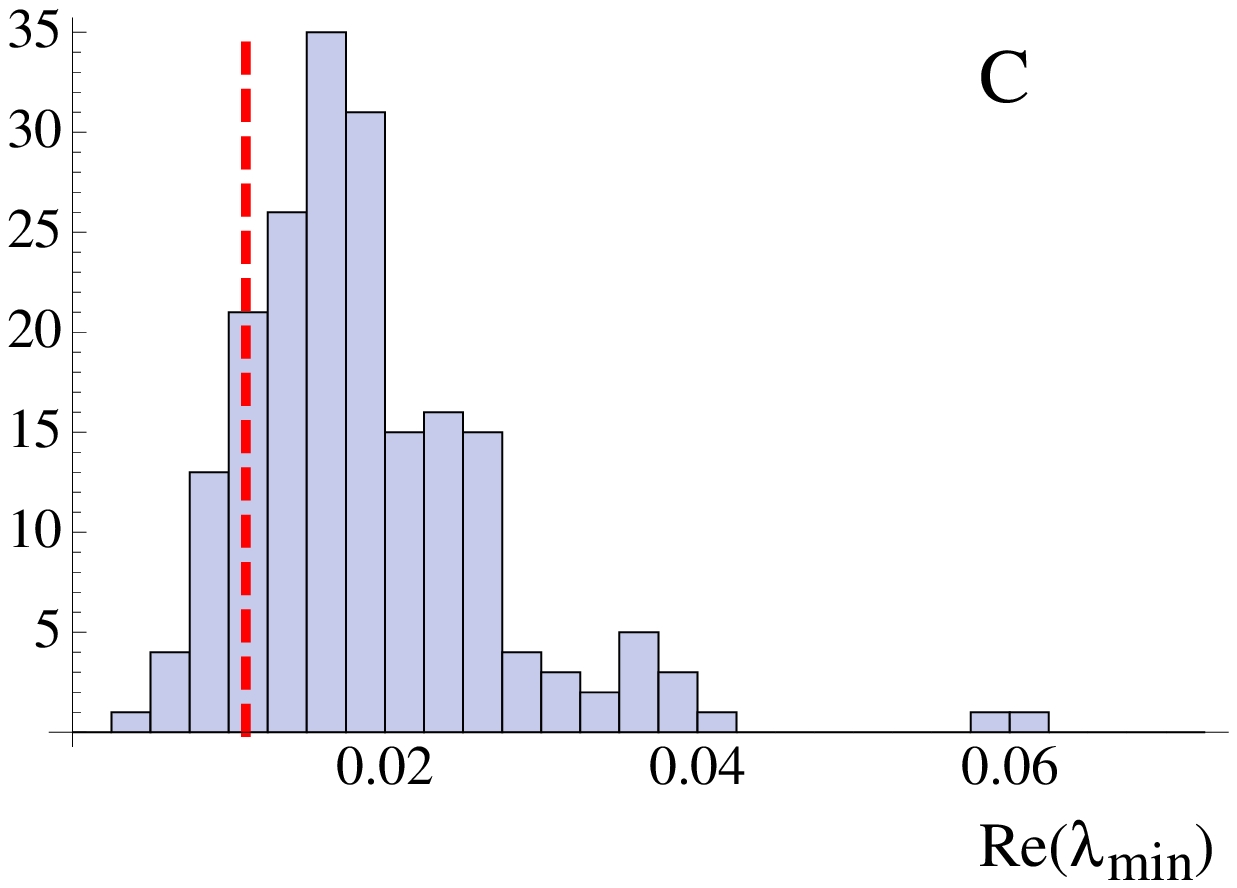}\hspace*{5mm}
\includegraphics*[width=7cm,clip]{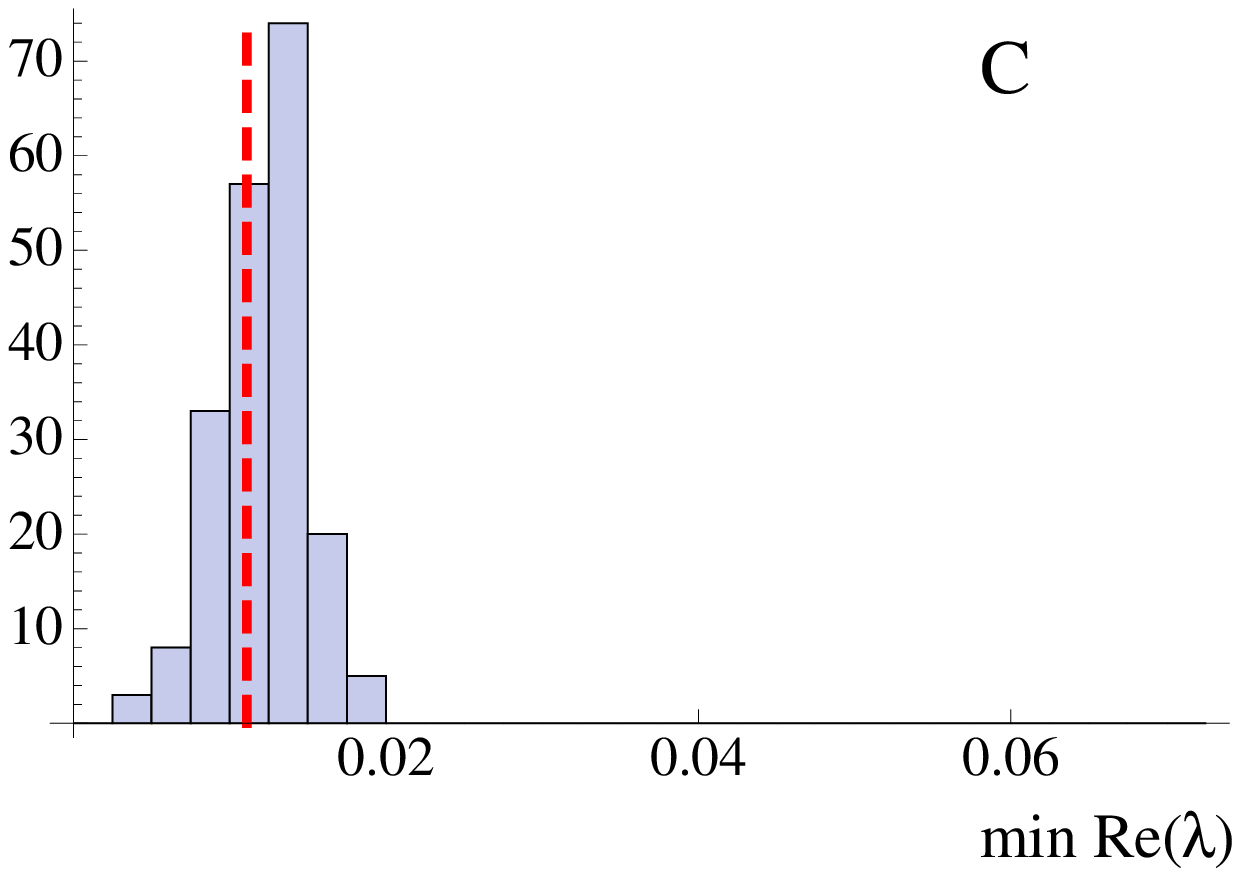}\\
\end{center}
\caption{\label{fig:eigenvalues_histogram} Histograms for the smallest values of
real $\lambda$ (left) and the smallest values of $\mathrm{Re}({\lambda})$
(right) for parameter sets A-C (from top to bottom). The measured AWI-mass (in
lattice units) is indicated by the vertical, dashed line.}
\end{figure}

Figure \ref{fig:eigenvalues_histogram} shows histograms for the smallest values
of purely real $\lambda$ and for the minimal $\mathrm{Re}(\lambda)$ for all
three parameter sets. Both types of histograms give  an indication on the
permissible values of the smallest quark mass we may obtain for that action,
lattice spacing and lattice size. Concerning exceptional configurations, we find
a mass gap indicating  that we are in a safe region of parameter values.

Several observations can be made from the eigenvalue distributions. Low lying
eigenvalues are depleted as expected for dynamical fermions due to the effect of
the determinant in the measure. The boundary close to the circular shape is
rather sharp towards larger values of $|\lambda-1|$. This allows to simulate
smaller pion  masses on coarse lattices.

The distribution density towards the inner region is, for a given scale
parameter, narrower than that for the Wilson action but not as close to the
boundary as for quenched simulations with \DCI \cite{GaHiLa00}. 

In the quenched simulation hypercubic smearing was used whereas for the
dynamical simulation we apply stout smearing. This latter type of smearing has a
weaker smoothing effect than the hypercubic type. We could have applied several
subsequent stout smearing steps instead, but we did not want to change the
effective action in the middle of our runs. Also, for the quenched ensembles we
optimized the action parameters for each value of $\beta_1$. In the dynamical
simulation we stayed with the same parameterization of the \DCI (except for the
bare ``mass'' parameter $m_0$) in order to be able to qualitatively compare
different runs.

\begin{figure}[tp]
\begin{center}
\includegraphics*[width=7cm,clip]{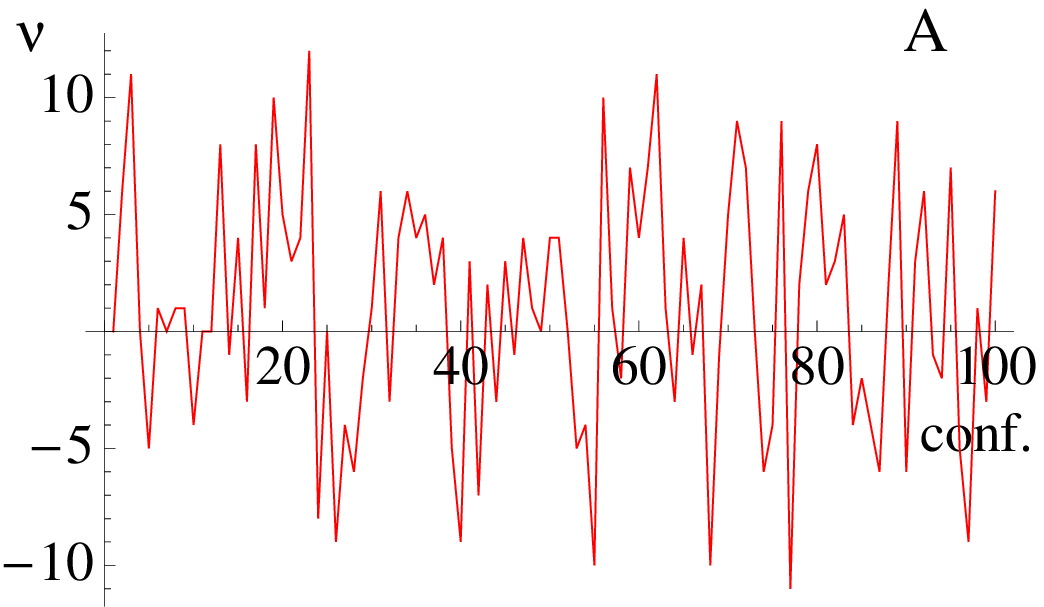}    \hspace{5mm}
\includegraphics*[width=7cm,clip]{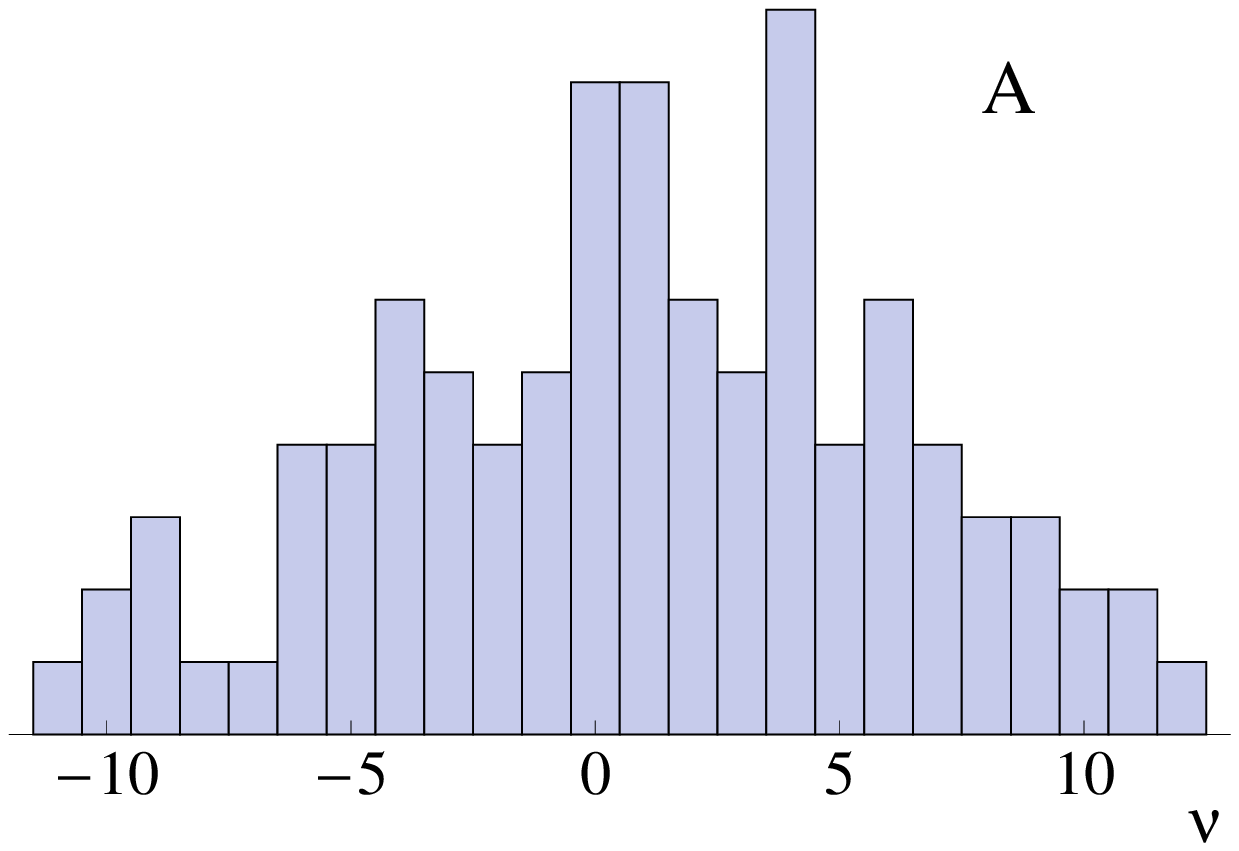}  \\
\includegraphics*[width=7cm,clip]{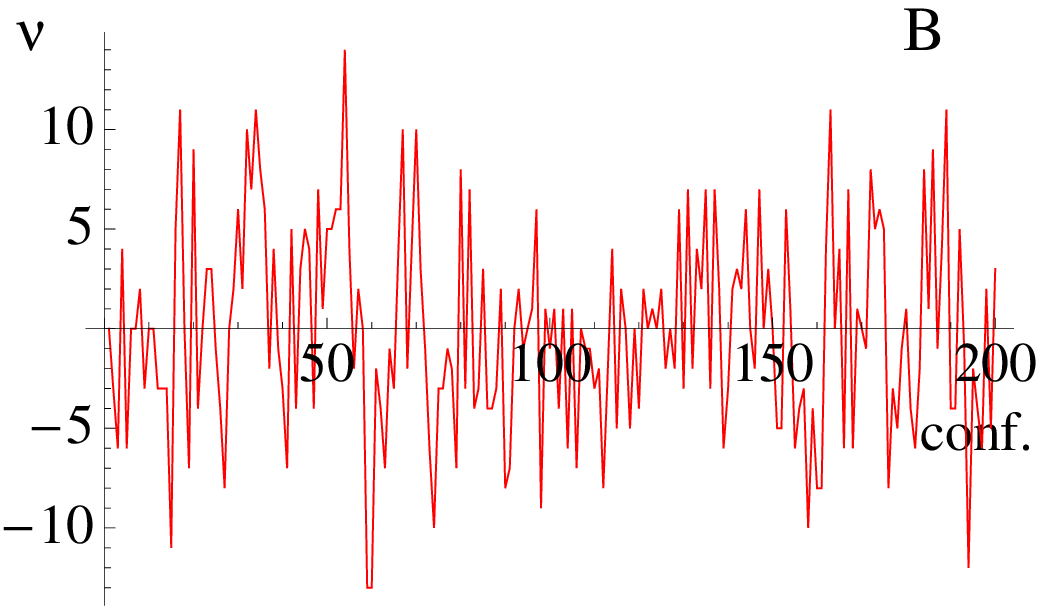}    \hspace{5mm}
\includegraphics*[width=7cm,clip]{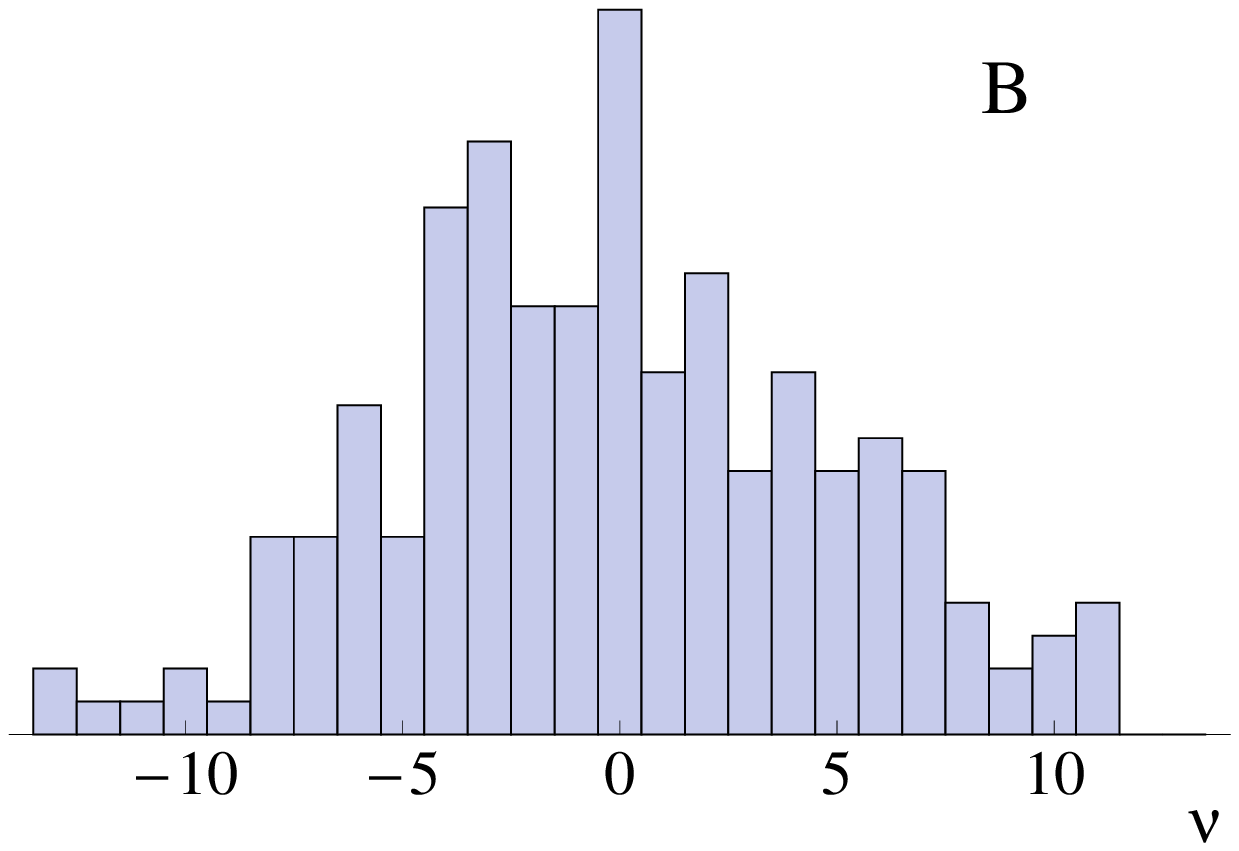}  \\
\includegraphics*[width=7cm,clip]{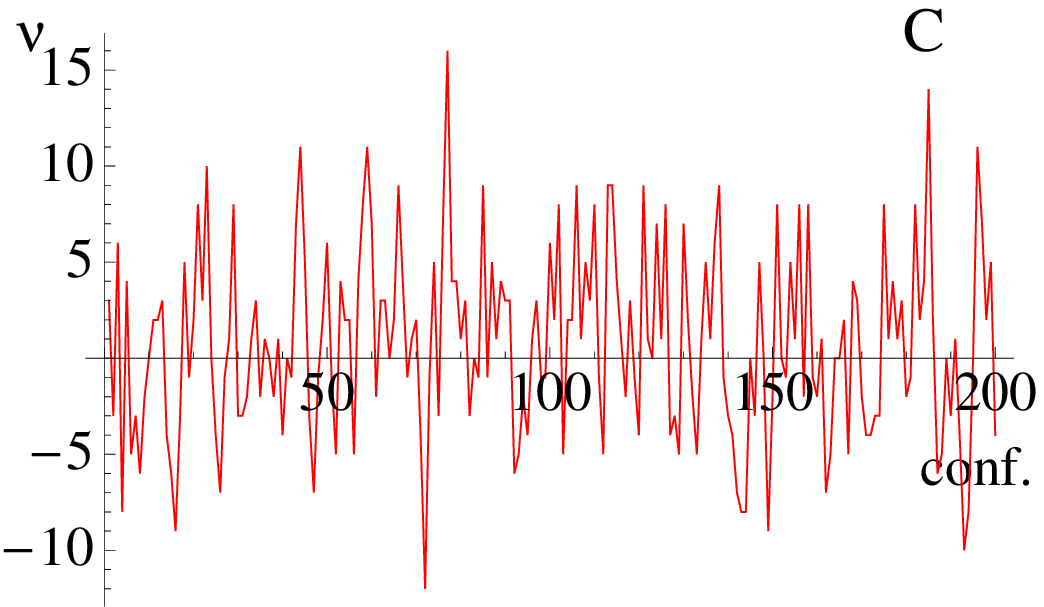}   \hspace{5mm}
\includegraphics*[width=7cm,clip]{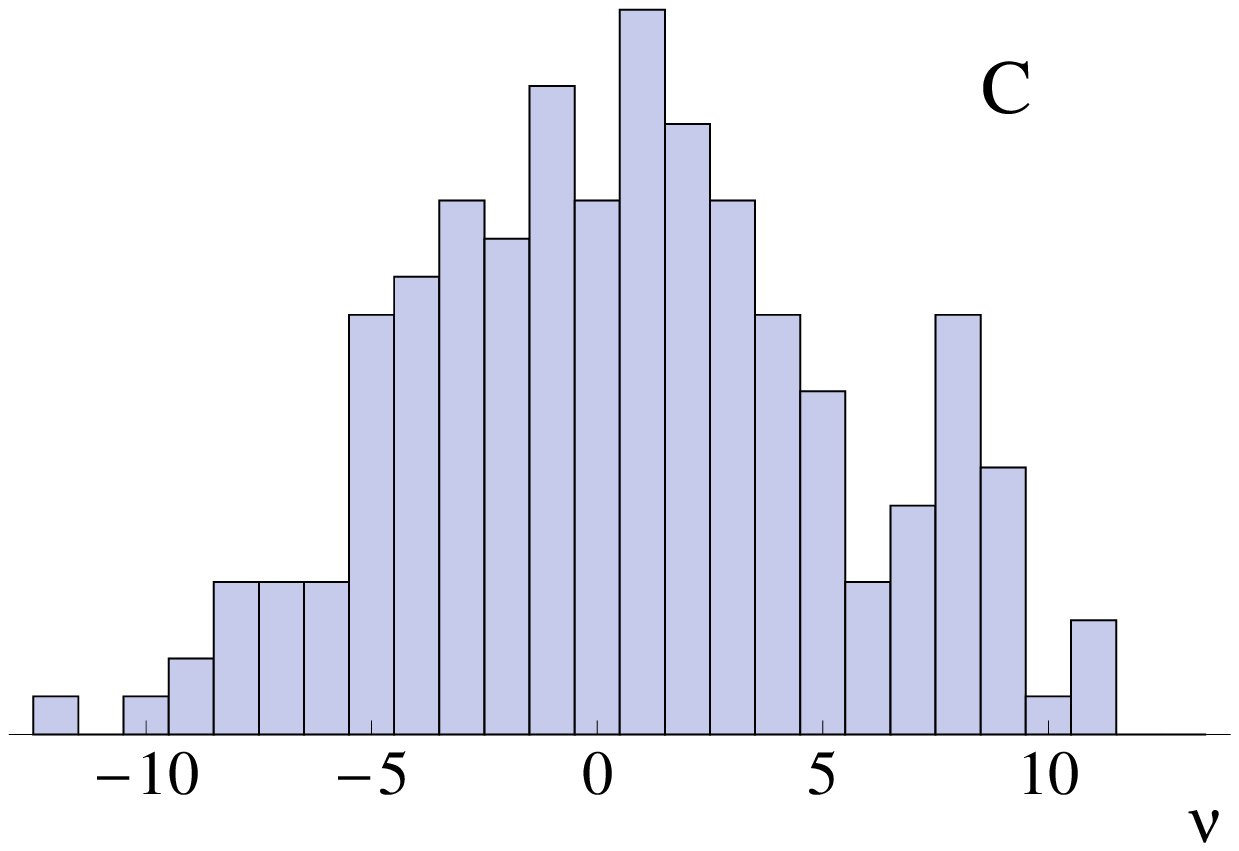} \\
\end{center}
\caption{\label{fig:topologytunneling}
History of the topology sector $\nu$ and corresponding distribution histogram 
for parameter sets A-C (from top to bottom)}
\end{figure}

The number of exactly real modes $\nu$, counted according to their chirality
$\langle \psi|\gamma_5|\psi\rangle$, may be related to the topological charge
via the Atiyah-Singer index theorem \cite{AtSi71}. Although we cannot
exclude that we miss some of the inner real modes (cf.,
Fig.~\ref{fig:eigenvalues_histogram}), we still get some information on the
tunneling between topological sectors from this quantity. Figure
\ref{fig:topologytunneling} demonstrates frequent tunneling and consistency with
a Gaussian-like shape of the distribution.

\section{Results for low energy parameters}\label{sec:results1}

\subsection{Setting the scale}

For the determination of the lattice spacing we used the Sommer parameter
\cite{So94}, determined by the lattice potential, which was derived from Wilson
loops $W(r,t)$. For improving the signal the gauge configurations have been
smeared with hypercubic blocking \cite{HaKn01} with parameter values $a_1=0.75$,
$a_2=0.6$ and $a_3=0.3$.

\begin{figure}[tb]
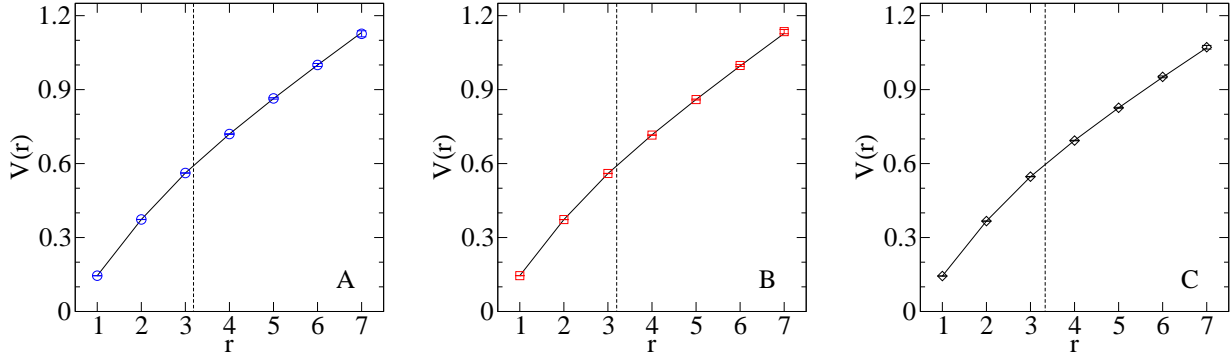
\begin{center}
\includegraphics[width=5cm,clip]{./figs/pot_v1c_A.eps}\hspace*{5mm}
\includegraphics[width=5cm,clip]{./figs/pot_v1c_B.eps}\hspace*{5mm}
\includegraphics[width=5cm,clip]{./figs/pot_v1c_C.eps}\\
\caption{\label{fig:pot_A} Fits to the potential in the range $1\le r \le 7$
(symbols represent data points, whereas full black lines are fits to these
points). The dashed black line in each plot indicates the distance $r$ in
lattice units where Eq.~\eqref{eq:1.65} holds.}
\end{center}\end{figure}

We have extracted the potential $V(r)$ for each value of $r$ from linear fits to
$\ln W(r,t)$ in the range $4\le t \le 7$. The potential was then fitted in the
range $1\le r \le 7$ to
\be
V(r) = A + \frac{B}{r} + \sigma\,r + C\, \Delta V(r)
\quad
\textrm{with}
\quad
 \Delta V(r)\equiv \left[\frac{1}{\VEC r}\right]-\frac{1}{r}
\ee
(all quantities given in lattice units). The perturbative lattice Coulomb
potential $[1/\VEC r]$ serves as a correction to the continuum Coulomb potential
as discussed in \cite{LaRe82,Mi85,EdHeKl98,AlBoBo02}. It has been used in the
form corrected for hypercubic blocking \cite{HaHoKn01,GaHoSc02a},
\be
\left[\frac{1}{\VEC r}\right]=\pi\int_{-\pi}^\pi \frac{d^3 k}{(2\pi)^3}
\frac{\cos(\VEC k\cdot \VEC r) \cdot S_\srm{HYP}(\VEC k)}{4\sum_{i=1}^3 \sin^2(k_i/2)}\ .
\ee
The smearing factor $S_\srm{HYP}(\VEC k)$ is detailed in \cite{HaHoKn01}. The
correction term allows for a perfect fit, even including the $r=1$ value, see
Fig.~\ref{fig:pot_A}. Actually, as observed by other authors, the result lies
very close to what one gets when fitting only the continuum shape of the
potential to a restricted range $2\le r \le 7$.

From the resulting potential without the correction term $\Delta V$ and the
condition
\be\label{eq:1.65}
r^2 \left.\frac{dV(r)}{dr}\right\vert_{r=r_0} = 1.65\ ,
\ee
we obtain the Sommer parameter in lattice units,
\be
r_0=\sqrt{\frac{1.65+B}{\sigma}}=\frac{r_{0,\text{exp}}}{a}\ .
\ee
The lattice spacing is thus given by $a=r_{0,\text{exp}}/r_0$. Using
$r_{0,\text{exp}}=0.48$ fm, our values for the lattice spacing are given in
Table \ref{tab:latt_a}.

\begin{table}\begin{center}
\begin{tabular}{ccccc}
\hline\hline
Run~ & $a$ [fm] & $a/r_{0,\text{exp}}$ & $a\,m_\srm{AWI}$ & ~$m_\srm{AWI}$ [MeV]\\
\hline
A & ~$0.1507(17)$~ & ~$0.3139(35)$~ & ~$0.0327(3)$~ & $42.8(4)$ \\
B & ~$0.1500(12)$~ & ~$0.3126(24)$~ & ~$0.0259(2)$~ & $34.1(2)$ \\
C & ~$0.1440(12)$~ & ~$0.3000(24)$~ & ~$0.0111(2)$~ & $15.3(3)$ \\
\hline\hline
\end{tabular}
\caption{Lattice spacing as defined via the Sommer parameter and AWI-mass in
lattice units and in physical units via that scale setting. \label{tab:latt_a}}
\end{center}\end{table}

The physical value of $r_{0,\text{exp}}$ for our situation  (two mass degenerate
quarks) is not accessible. Often the scale is set by extrapolating the measured
values of the lattice spacing to vanishing quark mass and using the 
extrapolated value for all mass values \cite{Ao01} (mass independent scheme). In
the present state of our simulations we have only one mass value for each gauge
coupling. We therefore rely on the mass dependent definition, which differs by
$\mathcal{O}(a)$ corrections. We also could use the nucleon mass to set the
scale.  In some of the mass plots shown below we therefore plot the masses in
units of the nucleon mass.

\subsection{The axial Ward identity mass}\label{sec:awimass}

\begin{figure}
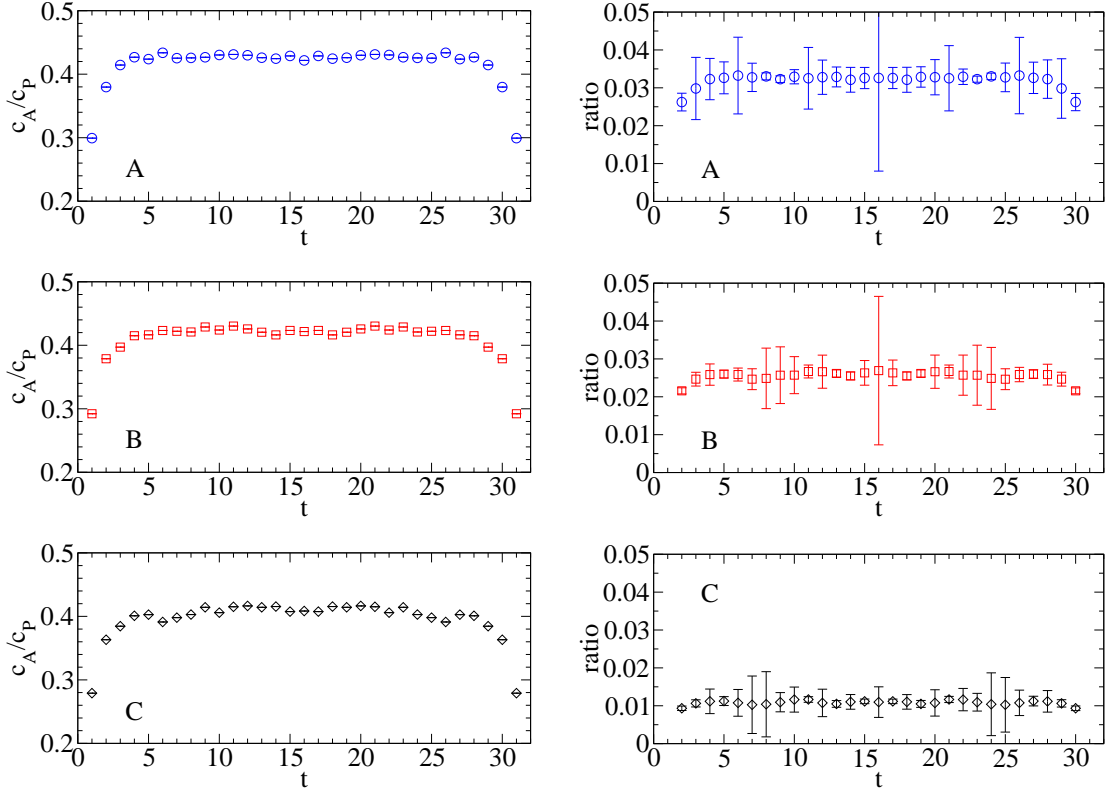
\begin{center}
\includegraphics[width=7cm,clip]{./figs/ca_cp_ratio_A.eps}\hspace*{5mm}
\includegraphics[width=7cm,clip]{./figs/awi_mass_A.eps}\\[3mm]
\includegraphics[width=7cm,clip]{./figs/ca_cp_ratio_B.eps}\hspace*{5mm}
\includegraphics[width=7cm,clip]{./figs/awi_mass_B.eps}\\[3mm]
\includegraphics[width=7cm,clip]{./figs/ca_cp_ratio_C.eps}\hspace*{5mm}
\includegraphics[width=7cm,clip]{./figs/awi_mass_C.eps}
\caption{L.h.s.: Ratio $c_A/c_P$ for each run. R.h.s.: The AWI-mass ratios from
Eq.~\eqref{eq:awimass} (runs A, B and C from top to bottom).
\label{fig:awimass}\label{fig:ca_cp}}
\end{center}\end{figure}

Another important observable in lattice QCD calculations is the 
(unrenormalized) quark mass from the axial Ward identity and the PCAC
relation, the so-called AWI-mass  (or PCAC-mass). Therefore, we compute the
ratio
\be \label{eq:awimass}
m_\srm{AWI} = \frac12\, \frac{c_A} {c_P}\, \frac{\la \partial_t A_4(\vec{p}=0,t)\, P(0)
\ra} {\la P(\vec{p}=0,t)\, P(0) \ra}\ .
\ee
Both interpolators 
\be
A_4 = \ov{d}\, \g_4\, \g_5\, u\ , \quad P = \ov{d}\, \g_5\, u\ ,
\ee
couple to the pseudoscalar meson channel (here, time is direction 4). 
In relating the lattice measurements
to the \MSbar-scheme, these operators are usually defined with point-like quark
sources. The normalization factors $c_A$, $c_P$ relate the smeared source
lattice operators to the point source lattice operators,
\be\label{eq:c_x}
c_X = \frac{\big\la X^{(p)}(t)\, P^{(p)}(0)\big\ra}
{\big\la X^{(s)}(t)\, P^{(p)}(0)\big\ra}\ ,
\ee
where the upper index ($s$) or ($p$) indicates smeared or point sources,
respectively, and $X$ refers to $A_4$ or $P$. The ratio $c_A/c_P$ is read off
from the plateau range as exhibited in Fig.~\ref{fig:ca_cp} for the case of the
wide sources.

For the ratio in Eq.~\eqref{eq:awimass} we need derivatives of the correlator
with respect to $t$. We obtain the numerical derivatives from local 3-point fits
to the expected  $\cosh$-behavior of the correlator, involving values at
$(t-1,t,t+1)$.

In Fig.~\ref{fig:awimass} we show the AWI-mass ratio Eq.~\eqref{eq:awimass}
vs.~$t$ and give the corresponding numbers in Table \ref{tab:latt_a}. The values
are symmetrized with regard to  $T/2$ and the error is estimated by single
elimination jackknife. To obtain the final value for $m_\srm{AWI}$, the ratio
was averaged from $t=4,\ldots,16$, weighted according to the statistical errors.

To relate this lattice value of $m_\srm{AWI}$ to an \MSbar-value (e.g., at
$\mu$=2 GeV)  we still need to extrapolate to the chiral limit and compute the
corresponding renormalization constants of the axial and the pseudoscalar
operators,
\be
m_\srm{\MSbar} = m_\srm{AWI} Z_A/Z_P\;.
\ee

\begin{figure}\begin{center}
\includegraphics[width=7cm,clip]{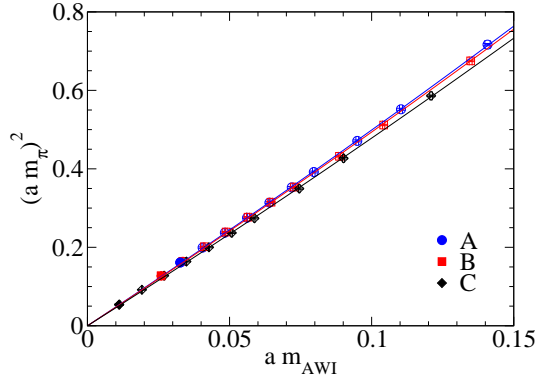}
\caption{Gell-Mann-Oakes-Renner plot for the three runs. Full symbols represent the fully
dynamical points whereas open symbols are data points for which $m_\srm{val}>m_\srm{sea}$.
The curves represent fits to $a\,m+b\,m^2$.}
\label{fig:gmor}
\end{center}\end{figure}

The Gell-Mann-Oakes-Renner (GMOR)  relation establishes (in leading
order in the quark mass) the connection between the pion mass $m_\pi$ and the
quark mass $m$, with $F_\pi$ and $\Sigma$ denoting the pion decay constant and
the chiral condensate,  respectively,
\be
F_\pi^2\,m_\pi^2 = -2\,m\,\Sigma\ .
\ee
From Fig.~\ref{fig:gmor} one sees that the expected linear dependence of
$m_\pi^2$ on $m_\srm{AWI}$ is nicely reproduced. In addition to the three fully
dynamical points we also show the partially quenched values, where the valence
quark mass is larger than the sea quark mass. These points, including the
partially quenched ones, are all compatible with a common behavior.

\subsection{Pion decay constant}\label{sec:pion_decay}

\begin{figure}\begin{center}
\includegraphics[width=7cm,clip]{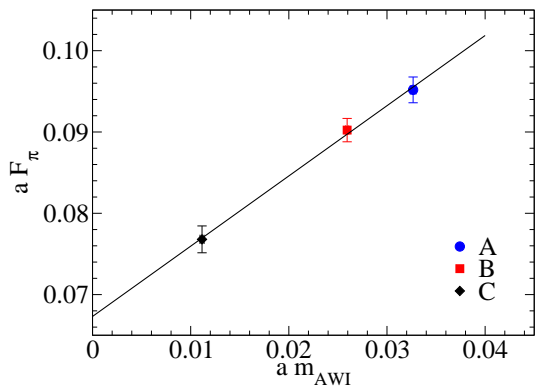}
\caption{$F_\pi$ in lattice units with a linear fit to the three dynamical
points. These values have not been corrected by multiplication with $Z_A$.}
\label{fig:pion_decay}
\end{center}\end{figure}

The pion decay constant $F_\pi$ can be extracted from the correlator $\la A_4\,
A_4\ra$ via
\be
c_A^2\,Z_A^2\, \la A_4 (\vec{p}=0,t)\, A_4(\vec{p}=0,0) \ra\  \stackrel{\text{large }t}
{\longrightarrow}\ m_\pi F_\pi^2 e^{-m_\pi t}\;.
\ee
Here, we also use the normalization factor $c_A$ from Eq.~\eqref{eq:c_x} in
order to remove the dependence on the quark smearing. The axial vector lattice
field operators have to be multiplied with normalization constants $Z_A$ in
order to ensure correct current conservation in the chiral limit. For the
quenched results these were determined for \DCI in \cite{GaHuLa05a}
resulting in values close to 1. In Fig.~\ref{fig:pion_decay} we plot $F_\pi$
vs.  the AWI-mass.

\section{Results for the hadron ground states}

\subsection{Spectrum analysis: Variational method}

Over the last two decades lattice QCD has turned into a powerful tool for
computing the mass spectrum of hadrons. Such a reproduction of experimental
evidence from an ab-initio calculation is a strong test for the correctness of
QCD. However, one mostly is restricted to the ground state masses, since excited
state contributions only appear as sub-leading terms in the Euclidean
correlators. Thus, a reliable separation of excited and ground states, but also
of different excited states themselves, is a rather challenging enterprise.

Nowadays several different approaches towards that goal are used in hadron
spectroscopy. One could do a brute-force least-squares fit to a finite sum of
exponentials, but this is known to give conclusive results only if high
statistics are available. Other methods are based on Bayesian fitting
\cite{Mi94,McMi95,AsHaNa01,LeClDa01,ChDoDr04}, subtractions \cite{GuPaSi04} or
evolutionary fitting methods \cite{HiLePe07a,HiLePe07}. Here, however, we use a
different state-of-the-art approach, namely the variational method
\cite{Mi85,LuWo90} which has been used quite extensively within the BGR
collaboration \cite{BuGaGl04a, BuGaGl05a, BuGaGl05b, BuGaGl05c, BuGaGl06a,
BuGaGl05d, BuGaGl06, BuGaGl06b, GaGlLa08}. For a recent review on results for
the variational method see \cite{Mo08}.

In the variational method a matrix built from different correlators is used. 
These correlators contain interpolators with different Dirac structure and
quarks smeared with different widths. Such a choice allows for a better overlap
of the interpolating fields with the physical states. Given a set of $N$ basis
interpolators $O_i, i=1,\ldots,N$, we compute a matrix of cross correlations,
\be
C_{ij}(t) = \langle\, O_i(t)\, \ov{O}_j(0)\, \rangle \ .
\ee
Considering the generalized eigenvalue problem, normalized at some time slice
$t_0<t$,
\be \label{eq:gevp}
C(t)\cdot \vec{v}_k = \lambda_k(t,t_0) \cdot C(t_0) \cdot \vec{v}_k\ ,
\ee
one can show along the lines of \cite{LuWo90,BlHiMe08}, that the eigenvalues
$\lambda_k$ behave as
\be \label{eq:eigenvalue}
\lambda_k(t,t_0) \propto e^{-(t-t_0)m_k}\left[ 1+\mathcal{O}(e^{-(t-t_0)\Delta m_k})
\right]\ .
\ee
In general, $\Delta m_k$ is the mass difference to the closest lying state. For
a more detailed discussion of the error terms see \cite{BlHiMe08}. Each of the
interpolators $O_i$ has the quantum numbers of the corresponding hadron channel 
and is projected to a certain spatial momentum, which is always zero in our
case. For all considered hadron channels we use $t_0=1$.

For a sufficiently large set of basis interpolators each eigenstate decays
exponentially with its energy according to Eq.~\eqref{eq:eigenvalue}. The
eigenstate with slowest decay (i.e., the largest eigenvalue) corresponds to the
ground state, the second largest to the first excited state, and so on. We now
can fit the states by stable two parameter fits of the eigenvalues in a range of
$t$-values where the correlator is dominated by a single exponential.  In order
to identify the corresponding range for the fit we analyze effective masses for
the eigenvalues,
\be
m_k^\text{(eff)}(t+1/2) = \ln \left( \frac{\lambda_k(t)}{\lambda_k(t+1)}\right) \ .
\ee
For sufficiently large values of $t$ the effective masses form plateaus, which
then give us the range for the fit.

Another important instrument to estimate the quality of the signal are the
eigenvectors $\vec{v}_k$ of Eq.\ (\ref{eq:gevp}), acting as fingerprints for
each state. The components also should show a plateau behavior with regard to
the correlation distance where the channel is dominated by a single state. Thus,
the fits of the eigenvalues should only be performed in a $t$-range where both
effective masses and eigenvectors show a reliable plateau.

\subsection{Jacobi smearing of quark sources}

Hadron correlation functions are built from quark propagators $D^{-1}$ acting on
some quark source $S$. In order to improve the signal and to extend the operator
basis we work with extended sources obtained by Jacobi smearing
\cite{Gu89,Be97}: A point-like source $S_0$ is smeared out by acting with a
smearing operator $M$,
\be
S = M S_0\ , \quad M=\sum_{n=0}^N\, (\kappa H)^n\ ,
\ee
where $H$ is a hopping term,
\be
H=\sum_{j=1}^3\, \left[ U_j(\vec{x},\,t)\,\delta_{\vec{x}+\hat{j},\vec{y}} +
U_j^\dagger(\vec{x}-\hat{j},\,t)\,\delta_{\vec{x}-\hat{j},\vec{y}} \right]\;.
\ee
The smearing extends only over individual time slices, i.e., $t$ is fixed. The
parameters $\kappa$ (hopping parameter) and $N$ (number of smearing steps) are
tuned to get an approximately Gaussian shape of the quark source with a certain
width. We use the values for $\kappa$ and $N$ given in \cite{BuGaGl06} for the
$16^3\times 32$ lattice to obtain a narrow (index $n$) and a wide (index $w$)
source.

\subsection{Hadron interpolators}\label{sec_hadron_interpolators}

Working with the variational method one strives for a good basis of
interpolators $O_i$ from which one can obtain a combination coupling strongly to
the hadron of interest.  These interpolators should simultaneously be linearly
independent, as orthogonal as possible and sufficient to represent the physical
states reasonably well. Thus, the crucial point is the design of different
interpolators.

A complete list of our meson interpolators can be found in
Tab.~\ref{tab:mesons}. All considered interpolators represent isovector ($I=1$)
mesons.

\begin{table}[tb]\begin{center}
\begin{tabular}{lccl}
\hline\hline
Meson & ~$J^{PC}$~ & ~Number~ & Operator \\ \hline
Pseudoscalar& $0^{+-}$ &   1  &  $\ov{u}_n \gamma_5 d_n $ \\
	   &  &   2  &  $\ov{u}_n \gamma_5 d_w $ \\
	   &  &   3  &  $\ov{u}_w \gamma_5 d_w $ \\
	   &  &   4  &  $\ov{u}_n \gamma_t\gamma_5 d_n $ \\
	   &  &   5  &  $\ov{u}_n \gamma_t\gamma_5 d_w $ \\
	   &  &   6  &  $\ov{u}_w \gamma_t\gamma_5 d_w $ \\
\hline
Vector &$1^{--}$ &   1  &  $\ov{u}_n \gamma_k d_n $ \\
	   &  &  2  &  $\ov{u}_n \gamma_k d_w $ \\
	   &  &  3  &  $\ov{u}_w \gamma_k d_w $ \\
	   &  &  4  &  $\ov{u}_n \gamma_k\gamma_t d_n $ \\
	   &  &  5  &  $\ov{u}_n \gamma_k\gamma_t d_w $ \\
	   &  &  6  &  $\ov{u}_w \gamma_k\gamma_t d_w $ \\
\hline\hline
\end{tabular}
\caption{Meson interpolators used in this study. We use $\gamma_t=\gamma_4$,
i.e., the 4-direction  corresponds to the Euclidean time direction. The subscripts $n$ or $w$
denote the narrow or wide smeared quark source. \label{tab:mesons}}
\end{center}\end{table}

Interpolators for baryons are slightly more complicated since there are three
quarks involved. The general form of a local interpolator for the nucleon  is
given by
\be\label{eq:nucleon_def}
O_N = \epsilon_{abc}\, \Gamma_1\, u_a\, \big( u_b^T\, \Gamma_2\, d_c -
d_b^T\, \Gamma_2\, u_c \big)\ ,
\ee
where $a,b,c$ are color indices and $\Gamma_1,\Gamma_2$ are combinations of
$\gamma$-matrices. In Tab.~\ref{tab:nucleon} the different possibilities  are
listed.

The delta baryon has a simpler structure, since there only one Dirac structure
is analyzed. Its interpolator has the following form,
\be\label{eq:delta_def}
O_{\Delta,k} = \epsilon_{abc}\, u_a\, \big(u_b^T\, C\, \g_k\, u_c \big)\ ,
\quad k=1,\,2,\,3\ .
\ee
We project this to spin $\frac{3}{2}$ and average the correlators as discussed
in  \cite{BuGaGl06b}.

We introduce a short-hand notation for the different baryon interpolators. We
denote them by $s_1(s_2s_3)$, where $s_i$ represents the smearing type of quark
$i$; e.g., in $n(ww)$ the first quark has a narrow smearing, the second and
third have wide smearings. The interpolators for the baryons studied here can be
found in Tables \ref{tab:nucleon} and \ref{tab:delta}. All baryon correlators
are projected to definite parity.

\begin{table}[tb]
\begin{center}
\begin{tabular}{cccc} 
\hline\hline
~$\Gamma_1$~ & ~$\Gamma_2$~ & ~Number~ & ~Smearing~ \\
\hline
$\mathds{1}$ & $C\g_5$    &1 & $n(nn)$ \\
 & &2 & $n(nw)$ \\
 & &3 & $n(wn)$ \\
 & &4 & $n(ww)$ \\
 & &5 & $w(wn)$ \\
 & &6 & $w(ww)$ \\ \hline
$i\mathds{1}$ & ~$C\g_4\g_5$~ & 13 & $n(nn)$ \\
 & & 14 & $n(nw)$ \\
 & & 15 & $n(wn)$ \\
 & & 16 & $n(ww)$ \\
 & & 17 & $w(wn)$ \\
 & & 18 & $w(ww)$ \\
\hline\hline
\end{tabular}
\caption{Nucleon $I(J^P)=\ot\left(\ot^+\right)$ interpolators. The reference
numbers of the interpolators are chosen to be consistent with earlier
publications \cite{BuGaGl06,BuGaGl06b}. \label{tab:nucleon}}
\end{center}
\end{table}

\begin{table}[tb]
\begin{center}
\begin{tabular}{cc} \hline\hline
~Number~ & ~Smearing~ \\
\hline
1 & $n(nn)$ \\
2 & $n(nw)$ \\
3 & $n(wn)$ \\
4 & $n(ww)$ \\
5 & $w(wn)$ \\
6 & $w(ww)$ \\
\hline\hline
\end{tabular}
\caption{Delta baryon $I(J^P)=\frac{3}{2}\left(\frac{3}{2}^+\right)$
interpolators. The reference numbers of the interpolators are chosen to be
consistent with earlier publications \cite{BuGaGl06,BuGaGl06b}.
\label{tab:delta}}
\end{center}
\end{table}

For subsequent configurations the quark sources (and thus the hadron
interpolators) are placed at alternating positions $d_i=(t,\vec x)$ with
\bey
\vec{d}_1 &=& (\ \, 0,\ \, 0,\ \, 0,\ \, 0)\ ,  \nonumber \\
\vec{d}_2 &=& (16,\ \, 0,\ \, 0,\ \, 0)\ ,   \nonumber \\
\vec{d}_3 &=& (\ \, 0,\ \, 8,\ \, 8,\ \, 8)\ ,  \nonumber \\
\vec{d}_4 &=& (16,\ \, 8,\ \, 8,\ \, 8)\ ,
\eey
in order to get better statistical decorrelation of the data. All hadron
interpolators are projected to vanishing spatial momentum.


\subsection{Effective masses and fit ranges}

Only a posteriori one can judge on the amount of independence of the
interpolators used. Including too many interpolators in the correlation matrix
increases the statistical noise in the diagonalization. Our aim is to get the
best signal in each channel and this is obtained by having the best plateaus in
the effective masses. We therefore, after studying the quality of results with
different subsets of interpolators, decided on as few of them as seemed
sufficient for a stable signal. For example, for the positive parity nucleon we
only included the interpolators $4-6$ and  $16-18$. The optimal selection may
be different when we study higher excitations and may include differently
smeared quark sources.

\subsection{The meson sector}

Since we simulate two mass-degenerate light quarks, interpolators of the form
$\ov{u}_n\Gamma d_w$ and $\ov{u}_w\Gamma d_n$ are identical. 
Tab.~\ref{tab:mesons} lists the  interpolators used. Due to the two different
possibilities for $\Gamma$ we have for the pseudoscalar and vector particle six
interpolators at hand.  Only a subset of these is used for the final analysis.

We restricted ourselves to fit only plateaus with three or more consecutive
points. In addition to that we started fits only at points for which $t-t_0\ge
2$. Table \ref{tab:meson_details} gives the information on the interpolators and
the fit ranges  used in the final analysis.

\begin{table}\begin{center}
\begin{tabular}{lcccc}
\hline\hline
Meson & ~Interpolator(s)~ & ~Run~ & ~Fit range~ & ~Mass [MeV]~  \\
\hline
Pseudoscalar & 3 & A  & 4--15 & 526(7)\\
 & & B & 4--15 & 469(4)\\
 & & C & 5--15 & 318(5)\\
\hline
Vector & 4,5,6 & A & 3--10 & 922(17)\\
 & & B & 3--13 & 897(13)\\
 & & C & 4--9 & 810(28)\\
\hline
\hline
\end{tabular}
\caption{Here we show the interpolators entering the final analysis and the best
fit ranges for the different runs. We also give the resulting mass values using
the lattice spacing given in Table\ \ref{tab:latt_a}.} 
\label{tab:meson_details}
\end{center}\end{table}

\subsubsection{The pseudoscalar meson}

\begin{figure}
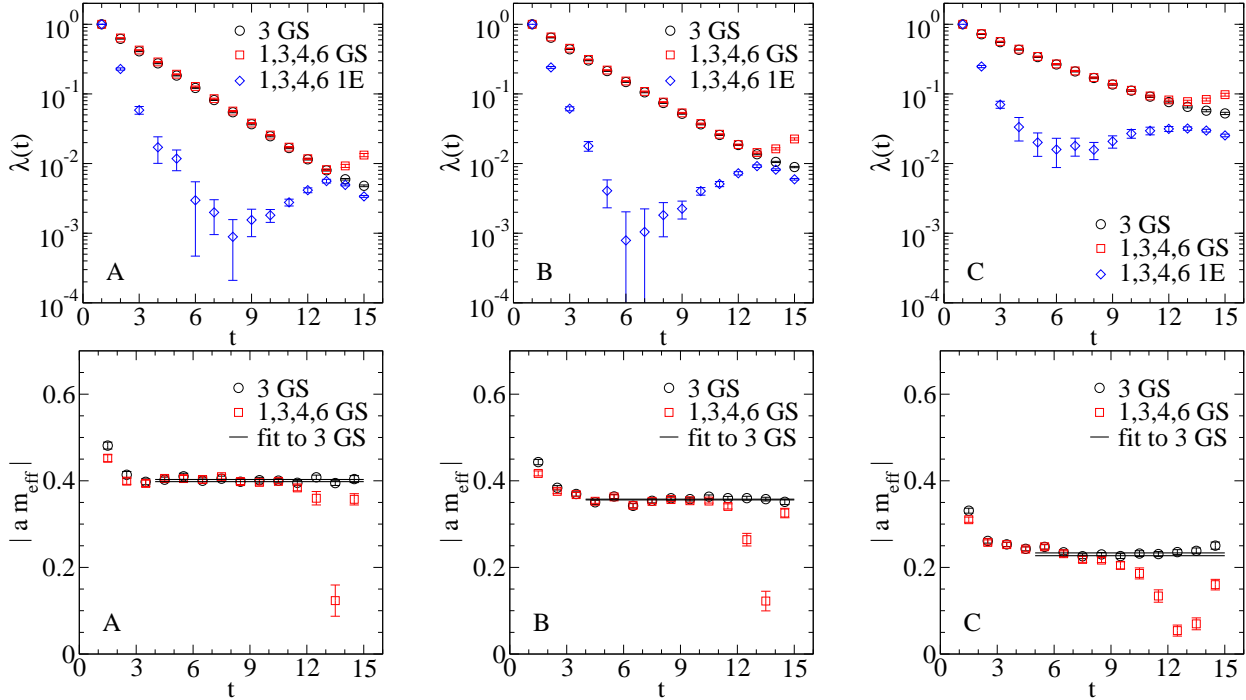
\begin{center}
\includegraphics[width=5cm,clip]{./figs/evals_pion_A.eps}\hfill
\includegraphics[width=5cm,clip]{./figs/evals_pion_B.eps}\hfill
\includegraphics[width=5cm,clip]{./figs/evals_pion_C.eps}\\
\includegraphics[width=5cm,clip]{./figs/m_eff_pion_A.eps}\hfill
\includegraphics[width=5cm,clip]{./figs/m_eff_pion_B.eps}\hfill
\includegraphics[width=5cm,clip]{./figs/m_eff_pion_C.eps}
\caption{In the first row the eigenvalues for the pseudoscalar channel
($J^{PC}=0^{-+}$) are shown (runs A, B, C from left to right). In each plot we
show data for two different sets of interpolators: Circles represent the ground
state  using interpolator $3$, squares and diamonds show the ground state (GS)
and the first excited state (1E) using the interpolator set $1,3,4,6$,
respectively (numbers according to Tab.~\ref{tab:mesons}). In the second row the
absolute value of the corresponding effective masses (in lattice units) of the
ground states are plotted as a function of $t$. The horizontal line indicates
the fit range and mass value obtained by the fit of the ground state eigenvalue
of interpolator $3$ over the specified range. }
\label{fig:pion}
\end{center}\end{figure}

Let us start our discussion with the particle where the best signal can be
extracted, the pseudoscalar meson ($J^{PC}=0^{-+}$). For the determination of
the ground state we used one interpolator (no.~3) and performed a cosh-fit.

Fig.~\ref{fig:pion} demonstrates a peculiarity of the generalized eigenvalue
problem, as it was observed already in, e.g.,  Refs. \cite{Li07a,GaGlLa08}. On
the periodically closed lattice mesons propagate forward and backward in time.
An interpolator which couples to a particular state at small $t$ will also
couple to the same, but backward running, state at high $t$. In the standard
eigenvalue problem, depending on the time extent and the masses of ground state
and excited state, above some value of $0 < t_1 \le n_t/2$ the backward running
ground state  will have a  larger eigenvalue than the first excited state. In
that region of $t$-values the second largest eigenvalue  increases towards
$n_t/2$.  In the generalized eigenvalue problem the eigenvalues are all
normalized to unity  at timeslice $t_0$. Thus the second largest eigenvalue
signal is shifted upwards and the upwards increasing eigenvalue discussed may
now even becomes at some value $t_2$ larger than the eigenvalue of the ground
state. This behavior is observed in Fig.~\ref{fig:pion} where we plot the first
two eigenvalues of the pseudoscalar state resulting from the generalized
eigenvalue problem analysis. For our choice of $t_0=1$ the (in time) backward
running ground state  becomes  the second largest eigenvalue near $t_1=6$ and
the largest eigenvalue near $t_2=13$. Near the crossing this leads to a
misidentification (the real ground state signal becomes the second largest
eigenvalue) which explains the bump in  the effective mass $\vert a
m_\srm{eff}\vert$.

These properties can be seen very nicely in the 2-dimensional model of
\cite{DaGa07}. This behavior of the eigenvalues is a fundamental feature of the
variational method; the signals of ground and excited states are disentangled up
to that point in time where these signals are crossing with the lightest
backward running state.  The larger the difference in the ground and excited
states, the earlier this crossing takes place. 

For simplicity, and since the results for the corresponding plateau regions
agree within errors, we choose the single correlator value where we find the
longest plateau. In Fig.~\ref{fig:pion} we compare the effective masses of the
ground states for two different choices of interpolators. One can clearly see
that the two sets of effective masses can be fitted reliably in an appropriate
region.

\subsubsection{The vector meson}

\begin{figure}
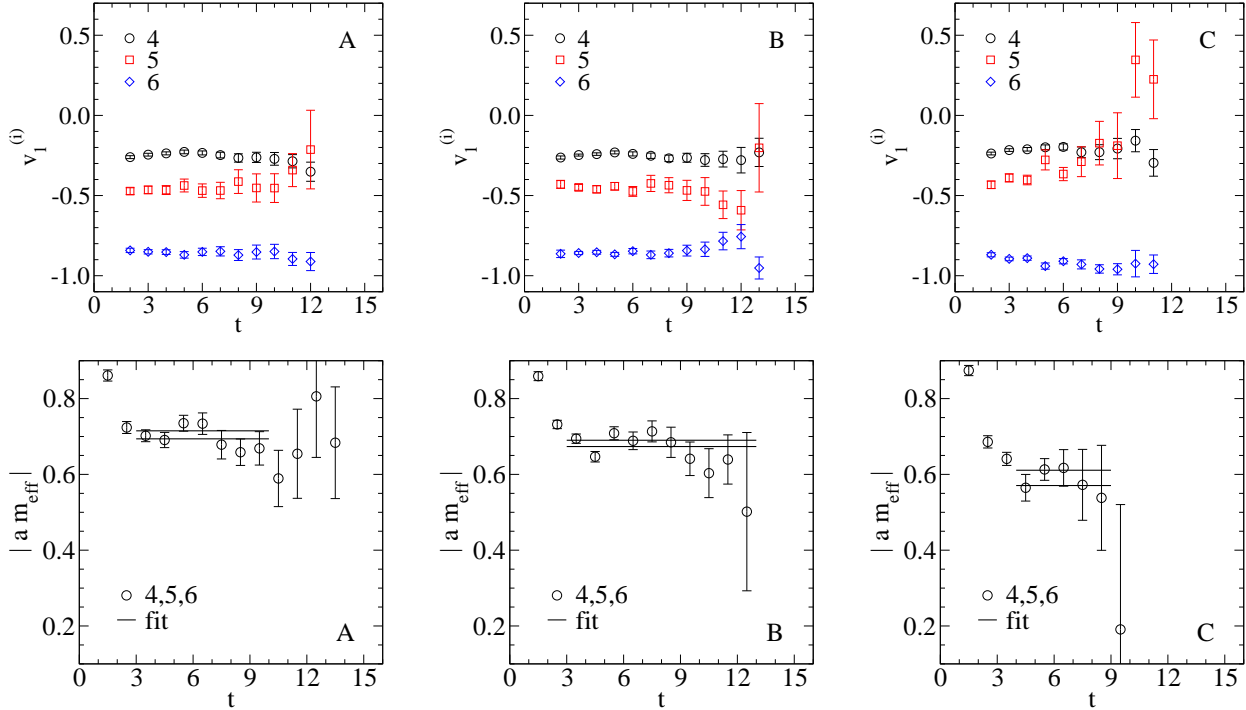
\begin{center}
\includegraphics[width=5cm,clip]{./figs/vector_vectors_1_A.eps}\hfill
\includegraphics[width=5cm,clip]{./figs/vector_vectors_1_B.eps}\hfill
\includegraphics[width=5cm,clip]{./figs/vector_vectors_1_C.eps}\\[3mm]
\includegraphics[width=5cm,clip]{./figs/m_eff_vector_A.eps}\hfill
\includegraphics[width=5cm,clip]{./figs/m_eff_vector_B_alternat.eps}\hfill
\includegraphics[width=5cm,clip]{./figs/m_eff_vector_C.eps}
\caption{In the first row the normalized eigenvector components of the ground
state $v_1^{(i)}, i=4,5,6$, of the vector meson ($J^{PC}=1^{--}$) are plotted
against the time distance $t$. From left to right we show runs A, B and C. In
the second row the effective masses of the vector meson (in lattice units) are
plotted as a function of $t$. The solid black lines indicate our fit range of
the fit to the corresponding leading eigenvalue and the upper and lower bound of
the extracted value. Also here, from left to right we show runs A,  B and C.
\label{fig:vector}}
\end{center}\end{figure}

In the case of the vector meson we can take the eigenvector components as a tool
to determine fit ranges (cf., Table~\ref{tab:meson_details}). The interpolators
we included in the correlation matrix are no.~$4,5,6$. In Fig.~\ref{fig:vector}
we plot the eigenvector components and effective mass of the ground state. One
can see from the plots that the quality of the data is sufficient to make a fit,
but it is not as good as for the pion.

\begin{figure}
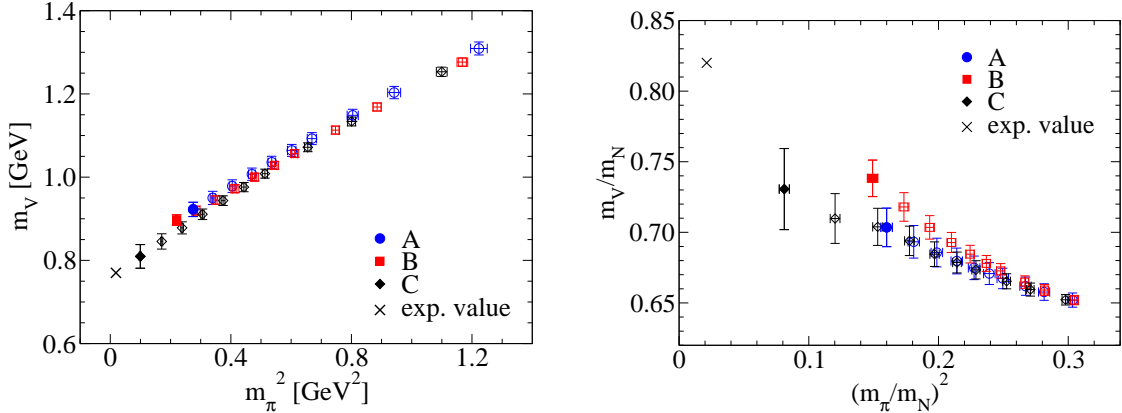
\begin{center}
\includegraphics[width=7cm,clip]{./figs/massplot_du_m3_alternat.eps}\hfil
\includegraphics[width=7cm,clip]{./figs/massplot_APE_nucleon_du_m3_alternat.eps}
\caption{L.h.s.: The vector meson mass $m_V$ is plotted against the pseudoscalar
mass $m_\pi^2$. R.h.s.: An APE plot (scaled by the nucleon mass $m_N$). Both
plots show the fully dynamical data (filled symbols) and the partially quenched
dynamical data (open symbols) for runs A, B, C. The physical point is marked
with a black cross. \label{fig:massplot_vector}}
\end{center}\end{figure}

On the l.h.s.~of Fig.~\ref{fig:massplot_vector} we plot the fitted mass against
$m_\pi^2$. The scale is set by the lattice spacing of Table \ref{tab:latt_a}.
Within error bars, all three runs agree nicely with each other and run C
extrapolates close to the experimental value. 

As discussed in Sect. \ref{sec:results1} we set the scale by assuming as Sommer
parameter value of 0.48 fm for all three runs. In the r.h.s.~of
Fig.~\ref{fig:massplot_vector} we use the scale of the nucleon mass instead.
Here the data of run B seem to be somewhat higher than runs A and C.

We emphasize that the physical $\rho$ is a resonance and that multiple lattice
volumes would be needed for a thorough analysis.

\subsection{The baryon sector}

In this presentation we restrict ourselves to baryons with positive parity. A
more detailed analysis, also including excited states, is in progress. The
definitions for the nucleon and delta baryons were given earlier in
Eqs.~\eqref{eq:nucleon_def} and \eqref{eq:delta_def}. Details of the
interpolators used can be found in Tables \ref{tab:nucleon} and \ref{tab:delta}.

\subsubsection{The nucleon}

\begin{figure}\begin{center}
\includegraphics[width=7cm,clip]{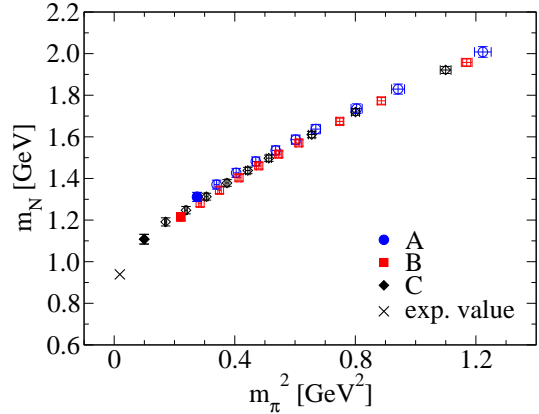}
\caption{The (positive parity) nucleon mass $m_N$ is plotted against the
pseudoscalar mass $m_\pi^2$ for the fully dynamical data (filled symbols) and
for the partially quenched dynamical data (open symbols) for runs A, B, C. The
experimental value is marked with a black cross. }
\label{fig:nucleon}
\end{center}\end{figure}

For the diagonalization process we used the interpolators $4,5,6,16,17,18$
according to Tab.~\ref{tab:nucleon}. The results are shown in
Fig.~\ref{fig:nucleon}. All of the data sets extrapolate towards the physical
point.

\subsubsection{The delta resonance}

\begin{figure}
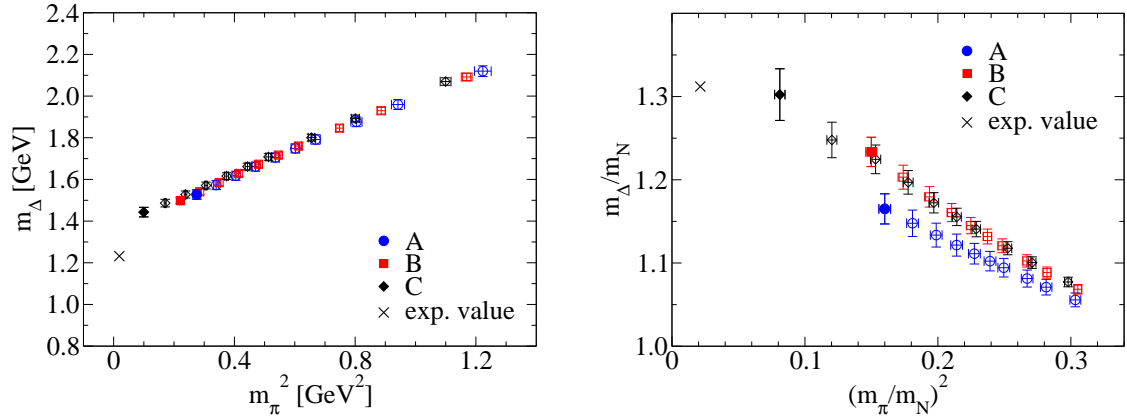
\begin{center}
\includegraphics[width=7cm,clip]{./figs/massplot_duu_b2.eps}\hfil
\includegraphics[width=7cm,clip]{./figs/massplot_APE_nucleon_duu_b2.eps}
\caption{L.h.s.: The (positive parity) delta baryon mass $m_\Delta$ is plotted
against the pseudoscalar mass $m_\pi^2$. R.h.s.: An APE plot (scaled by the
nucleon mass) for the dynamical runs A, B, C. Both plots show the fully
dynamical data (filled symbols) and the partially quenched dynamical data (open
symbols) for runs A, B, C. The physical point is marked with a black cross.
\label{fig:delta}}
\end{center}\end{figure}

For the $\Delta$ resonance we have a set of 6 different interpolators at hand
(see Tab.~\ref{tab:delta}) and in principle we can allow for $2^6-1=63$
combinations. All these combinations give rise to reasonable fit results. In the
end we used the combination $1,2,4,5,6$, see Fig.~\ref{fig:delta}. However, a
naive (linear  in $m_\pi^2$) extrapolation overestimates the physical value by
about 10\% - 15\%. On the r.h.s.~of Fig.~\ref{fig:delta} we show an APE plot,
scaled by the nucleon mass. One may argue that in this plot some finite size
artefacts cancel such that the extrapolation to the physical point is improved.

\begin{table}\begin{center}
\begin{tabular}{lcccc}
\hline\hline
Baryon & ~Interpolator(s)~ & ~Run~ & ~Fit range~ & ~Mass [MeV]  \\
\hline
Nucleon (pos.~parity)~ & 4,5,6,16,17,18 & A & 3--11 & 1311(22)\\
 & & B & 4--11 & 1215(18)\\
 & & C & 3--8 & 1108(23)\\
\hline
Delta (pos.~parity) & 1,2,4,5,6 & A & 3--6 & 1528(22)\\
 & & B & 3--6 & 1498(15)\\
 & & C & 3--6 & 1443(23)\\
\hline
\hline
\end{tabular}
\caption{Interpolators and fit ranges used for the baryon ground states. The
mass values are obtained using the lattice spacing given in  Table
\ref{tab:latt_a}. \label{tab:baryon_details}}
\end{center}\end{table}


\section{Summary and conclusions}

In this paper we presented first results from dynamical simulations with CI
fermions on lattices of size $16^3\times 32$ with spatial extent of 2.4 fm. 
After detailing the technical aspects of our simulation we showed that so-called
exceptional configurations are suppressed in simulations with CI fermions. This
enables us to simulate at pion masses of roughly $320$ MeV on rather coarse
lattices. We observe frequent tunneling between topological sectors and
reasonably small autocorrelation times.

As a first physical application we presented results for the pion decay constant
$F_\pi$ and for the ground state masses of selected mesons and baryons. While
scale setting remains an issue with dynamical simulations, the results from all
three runs are consistent and naive extrapolations of our data are also
consistent with experiment. Further simulations at different lattice spacings
and in larger volumes will be needed in order to control the effects of the
lattice discretization and to estimate the finite volume corrections, thereby
making closer contact with experimental results.

We are currently improving the basis for the variational method and
investigating the effects of quark and link smearing on the quality of excited
state signals, thus providing a systematic study of excited meson and baryon
states for a larger set of quantum numbers.


\begin{acknowledgments}

We are grateful to Meinulf G\"ockeler for discussions. We also thank T. Maurer
for help in an earlier stage of this work. The calculations have been performed
on the SGI Altix 4700 of the Leibniz-Rechenzentrum Munich and on local clusters
at ZID at the University of Graz. We thank these institutions for providing
support. M.L.~and D.M.~are supported by ``Fonds zur F{\"o}rderung der
wissenschaflichen Forschung in {\"O}sterreich'' (DK W1203-N08). C.H.~and
A.S.~acknowledge support by DFG and BMBF. The work has been supported by DFG
project SFB/TR-55.

\end{acknowledgments}


\appendix

\section{CI operator and L{\"u}scher-Weisz gauge action}

\subsection{The CI operator\label{app:DCI}}

Throughout the dynamical simulations we used the CI Dirac operator introduced in
\cite{Ga01a,GaHiLa00,GaGoRa01b}. The coefficients multiply terms of the action
according to the definition in
\be
D= m_0 \mathds{1} + \DCI\ , \ \DCI(n,m)= \sum_{i=1}^{16} c_{nm}^{(i)}(U)\; \Gamma_i\ ,
\ee
where the sum runs over all $16$ elements $\Gamma_i$ of the Clifford algebra. To
each element we assign a coefficient $c_{nm}^{(i)}$, consisting of sums of path
ordered products of the link variables $U$ which connect the lattice sites $n$
and $m$. Plugging this ansatz into the Ginsparg-Wilson equation leads to a set
of algebraic equations, which can be solved to obtain $\DCI$. Additional
restrictions come from the lattice symmetries and $\gamma_5$-hermiticity. The
solution can in principle be exact if one allows for an infinite number of
terms. For practical reasons the number of terms is finite and thus the solution
is a truncated series solution of the Ginsparg-Wilson relation.  In our
simulation paths up to length four are used, given in Table \ref{tab:dcicoeffs}.

\begin{table}[t]
\begin{center}
\begin{tabular}{cclccr}
\hline\hline
Coeff.~number & ~~Name~~ &~~~~~~Value & ~~Path shape~~ & ~$\gamma$~ & Multiplicity\cr
\hline
$ 1$ & $  s_1$  & $\wm 1.481599252    $ & $[\ ]       $ & $\unitmatrix$       & $  1$ \cr
$ 2$ & $  s_2$  & $-0.05218251439  $    & $[i]        $ & $\unitmatrix$       & $  8$ \cr
$ 3$ & $  s_3$  & $-0.01473643847  $    & $[i,j]	  $ & $\unitmatrix$       & $ 48$ \cr
$ 5$ & $  s_5$  & $-0.002186103421 $    & $[i,j,k]    $ & $\unitmatrix$	      & $192$ \cr
$ 6$ & $  s_6$  & $\wm 0.002133989696 $ & $[i,i,j]    $ & $\unitmatrix$	      & $ 96$ \cr
$ 8$ & $  s_8$  & $-0.003997001821 $    & $[i,j,-i]   $ & $\unitmatrix$	      & $ 48$ \cr
$10$ & $s_{10}$ & $-0.0004951673735$    & $[i,j,k,l]  $ & $\unitmatrix$	      & $384$ \cr
$11$ & $s_{11}$ & $-0.0009836500799$    & $[i,j,-i,k] $ & $\unitmatrix$	      & $384$ \cr
$13$ & $s_{13}$ & $\wm 0.007529838581 $ & $[i,j,-i,-j]$ & $\unitmatrix$	      & $ 48$ \cr
$14$ & $  v_1$  & $\wm 0.1972229309   $ & $[i]        $ & $\gamma_i$          & $  8$
\cr
$15$ & $  v_2$  & $\wm 0.008252157565 $ & $[i,j]	  $ & $\gamma_i$          & $ 96$
\cr
$17$ & $  v_4$  & $\wm 0.005113056314 $ & $[i,j,k]    $ & $\gamma_i$          & $384$
\cr
$18$ & $  v_5$  & $\wm 0.001736609425 $ & $[j,i,k]    $ & $\gamma_i$          & $192$
\cr
$32$ & $  t_1$  & $-0.08792744664  $    & $[i,j]	  $ & $\gamma_i \gamma_\nu$ & $ 48$
\cr
$33$ & $  t_2$  & $-0.002553055577 $    & $[i,j,k]    $ & $\gamma_i \gamma_j$ & $384$
\cr
$34$ & $  t_3$  & $\wm 0.002093792069 $ & $[i,k,j]    $ & $\gamma_i \gamma_j$ & $192$
\cr
$36$ & $  t_5$  & $-0.005567377075 $    & $[i,j,-i]   $ & $\gamma_i \gamma_j$ & $ 48$
\cr
$46$ & $t_{15}$ & $-0.003427310798 $    & $[j,i,-j,-i]$ & $\gamma_i \gamma_j$ & $ 48$
\cr
$51$ & $  p_1$  & $-0.008184103136 $    & $[i,j,k,l]  $ & $\gamma_5$          & $384$ \cr
\hline\hline
\end{tabular}
\end{center}
\caption{Coefficients for the CI fermion action used in this simulation. The
path shapes are given symbolically, e.g., $[i,j]$ stands for a path in
$i$-direction and then in $j$-direction ($i\neq j$). The $\gamma$-matrices (5-th
column) are also permuted as described in more detail in \cite{GaHiLa00}.}
\label{tab:dcicoeffs}
\end{table}

\subsection{The L{\"u}scher-Weisz gauge action\label{app:LW}}

The L{\"u}scher-Weisz gauge action \cite{LuWe85} is given by
\be
S_\srm{g} = -\beta_1\sum_\srm{pl}\frac{1}{3}\,\text{Re}\,\text{tr}\: U_\srm{pl}
      -\beta_2\sum_\srm{re}\frac{1}{3}\,\text{Re}\,\text{tr}\: U_\srm{re} 
    - \beta_3\sum_\srm{tb}\frac{1}{3}\,\text{Re}\,\text{tr}\: U_\srm{tb}\ ,
\ee
where $U_\srm{pl}$ is the usual Wilson plaquette, $U_\srm{re}$ is a planar
$(2\times1)$-plaquette and $U_\srm{tb}$ is a closed loop of length $6$ along the
edges of a $3$-cube (``twisted bent''). Here, $\beta_1$ is the independent gauge
coupling and the two other couplings are determined  from tadpole-improved
perturbation theory \cite{AlDiLe95}. With
\be \label{eq:ass_plaq}
u_0 = \left( \frac{1}{3}\,\text{Re}\,\text{Tr}\: \langle U_\srm{pl} \rangle
\right)^{1/4}\ ,\quad \alpha = -\frac{1}{3.06839} \log u_0^4\ ,
\ee
we get for $\beta_2, \beta_3$ the following expressions,
\be
\beta_2 = \frac{\beta_1}{20u_0^2}\, (1+0.4805\alpha)\ , \quad
\beta_3 = \frac{\beta_1}{u_0^2}\,  0.03325\alpha\ .
\ee
By $u_0$ in Eq.~\eqref{eq:ass_plaq} we denote the assumed plaquette,
$\text{Re}\,\text{Tr}\: \langle U_\srm{pl} \rangle$, thus the coefficients have
to be calculated self-consistently.




\begin{thebibliography}{78}
\expandafter\ifx\csname natexlab\endcsname\relax\def\natexlab#1{#1}\fi
\expandafter\ifx\csname bibnamefont\endcsname\relax
  \def\bibnamefont#1{#1}\fi
\expandafter\ifx\csname bibfnamefont\endcsname\relax
  \def\bibfnamefont#1{#1}\fi
\expandafter\ifx\csname citenamefont\endcsname\relax
  \def\citenamefont#1{#1}\fi
\expandafter\ifx\csname url\endcsname\relax
  \def\url#1{\texttt{#1}}\fi
\expandafter\ifx\csname urlprefix\endcsname\relax\def\urlprefix{URL }\fi
\providecommand{\bibinfo}[2]{#2}
\providecommand{\eprint}[2][]{\url{#2}}

\bibitem[{\citenamefont{Ginsparg and Wilson}(1982)}]{GiWi82}
\bibinfo{author}{\bibfnamefont{P.~H.} \bibnamefont{Ginsparg}} \bibnamefont{and}
  \bibinfo{author}{\bibfnamefont{K.~G.} \bibnamefont{Wilson}},
  \bibinfo{journal}{Phys. Rev. D} \textbf{\bibinfo{volume}{25}},
  \bibinfo{pages}{2649}
 (\bibinfo{year}{1982}).

\bibitem[{\citenamefont{L{\"u}scher}(1998)}]{Lu98}
\bibinfo{author}{\bibfnamefont{M.}~\bibnamefont{L{\"u}scher}},
  \bibinfo{journal}{Phys. Lett. B} \textbf{\bibinfo{volume}{428}},
  \bibinfo{pages}{342} (\bibinfo{year}{1998}), \eprint{hep-lat/9802011}.

\bibitem[{\citenamefont{Neuberger}(1998{\natexlab{a}})}]{Ne98a}
\bibinfo{author}{\bibfnamefont{H.}~\bibnamefont{Neuberger}},
  \bibinfo{journal}{Phys. Lett. B} \textbf{\bibinfo{volume}{417}},
  \bibinfo{pages}{141} (\bibinfo{year}{1998}{\natexlab{a}}),
\eprint{hep-lat/9707022}.

\bibitem[{\citenamefont{Neuberger}(1998{\natexlab{b}})}]{Ne98b}
\bibinfo{author}{\bibfnamefont{H.}~\bibnamefont{Neuberger}},
  \bibinfo{journal}{Phys. Lett. B} \textbf{\bibinfo{volume}{427}},
  \bibinfo{pages}{353} (\bibinfo{year}{1998}{\natexlab{b}}),
\eprint{hep-lat/9801031}.

\bibitem[{\citenamefont{Kaplan}(1992)}]{Ka92}
\bibinfo{author}{\bibfnamefont{D.~B.} \bibnamefont{Kaplan}},
  \bibinfo{journal}{Phys. Lett. B} \textbf{\bibinfo{volume}{288}},
  \bibinfo{pages}{342} (\bibinfo{year}{1992}),
\eprint{hep-lat/9206013}.

\bibitem[{\citenamefont{Furman and Shamir}(1995)}]{FuSh95}
\bibinfo{author}{\bibfnamefont{V.}~\bibnamefont{Furman}} \bibnamefont{and}
  \bibinfo{author}{\bibfnamefont{Y.}~\bibnamefont{Shamir}},
  \bibinfo{journal}{Nucl. Phys.} \textbf{\bibinfo{volume}{B439}},
  \bibinfo{pages}{54} (\bibinfo{year}{1995}), \eprint{hep-lat/9405004}.

\bibitem[{\citenamefont{Hasenfratz and Niedermayer}(1994)}]{HaNi94}
\bibinfo{author}{\bibfnamefont{P.}~\bibnamefont{Hasenfratz}} \bibnamefont{and}
  \bibinfo{author}{\bibfnamefont{F.}~\bibnamefont{Niedermayer}},
  \bibinfo{journal}{Nucl. Phys.} \textbf{\bibinfo{volume}{B414}},
  \bibinfo{pages}{785} (\bibinfo{year}{1994}), \eprint{hep-lat/9308004}.

\bibitem[{\citenamefont{Hasenfratz et~al.}(2005)\citenamefont{Hasenfratz,
  Hasenfratz, and Niedermayer}}]{HaHaNi05}
\bibinfo{author}{\bibfnamefont{A.}~\bibnamefont{Hasenfratz}},
  \bibinfo{author}{\bibfnamefont{P.}~\bibnamefont{Hasenfratz}},
  \bibnamefont{and}
  \bibinfo{author}{\bibfnamefont{F.}~\bibnamefont{Niedermayer}},
  \bibinfo{journal}{Phys. Rev. D} \textbf{\bibinfo{volume}{72}},
  \bibinfo{pages}{114508} (\bibinfo{year}{2005}), \eprint{hep-lat/0506024}.

\bibitem[{\citenamefont{Gattringer}(2001)}]{Ga01a}
\bibinfo{author}{\bibfnamefont{C.}~\bibnamefont{Gattringer}},
  \bibinfo{journal}{Phys. Rev. D} \textbf{\bibinfo{volume}{63}},
  \bibinfo{pages}{114501} (\bibinfo{year}{2001}), \eprint{hep-lat/0003005}.

\bibitem[{\citenamefont{Gattringer
  et~al.}(2001{\natexlab{a}})\citenamefont{Gattringer, Hip, and
  Lang}}]{GaHiLa00}
\bibinfo{author}{\bibfnamefont{C.}~\bibnamefont{Gattringer}},
  \bibinfo{author}{\bibfnamefont{I.}~\bibnamefont{Hip}}, \bibnamefont{and}
  \bibinfo{author}{\bibfnamefont{C.~B.} \bibnamefont{Lang}},
  \bibinfo{journal}{Nucl. Phys.} \textbf{\bibinfo{volume}{B597}},
  \bibinfo{pages}{451} (\bibinfo{year}{2001}{\natexlab{a}}),
  \eprint{hep-lat/0007042}.

\bibitem[{\citenamefont{Kaneko et~al.}(2006)\citenamefont{Kaneko, Aoki, Fukaya,
  Hashimoto, Ishikawa, Kanaya, Matsufuru, Okamoto, Okawa, Onogi
  et~al.}}]{KaAoFu06}
\bibinfo{author}{\bibfnamefont{T.}~\bibnamefont{Kaneko}},
  \bibinfo{author}{\bibfnamefont{S.}~\bibnamefont{Aoki}},
  \bibinfo{author}{\bibfnamefont{H.}~\bibnamefont{Fukaya}},
  \bibinfo{author}{\bibfnamefont{S.}~\bibnamefont{Hashimoto}},
  \bibinfo{author}{\bibfnamefont{K.-I.} \bibnamefont{Ishikawa}},
  \bibinfo{author}{\bibfnamefont{K.}~\bibnamefont{Kanaya}},
  \bibinfo{author}{\bibfnamefont{H.}~\bibnamefont{Matsufuru}},
  \bibinfo{author}{\bibfnamefont{M.}~\bibnamefont{Okamoto}},
  \bibinfo{author}{\bibfnamefont{M.}~\bibnamefont{Okawa}},
  \bibinfo{author}{\bibfnamefont{T.}~\bibnamefont{Onogi}},
  \bibnamefont{et~al.}, \bibinfo{journal}{PoS}
  \textbf{\bibinfo{volume}{LAT2006}}, \bibinfo{pages}{054}
  (\bibinfo{year}{2006}), \eprint{hep-lat/0610036}.

\bibitem[{\citenamefont{Matsufuru et~al.}(2006)\citenamefont{Matsufuru, Fukaya,
  Hashimoto, Kanaya, Kaneko, Ogawa, Okamoto, Onogi, and Yamada}}]{MaFuHa06}
\bibinfo{author}{\bibfnamefont{H.}~\bibnamefont{Matsufuru}},
  \bibinfo{author}{\bibfnamefont{H.}~\bibnamefont{Fukaya}},
  \bibinfo{author}{\bibfnamefont{S.}~\bibnamefont{Hashimoto}},
  \bibinfo{author}{\bibfnamefont{K.}~\bibnamefont{Kanaya}},
  \bibinfo{author}{\bibfnamefont{T.}~\bibnamefont{Kaneko}},
  \bibinfo{author}{\bibfnamefont{K.}~\bibnamefont{Ogawa}},
  \bibinfo{author}{\bibfnamefont{M.}~\bibnamefont{Okamoto}},
  \bibinfo{author}{\bibfnamefont{T.}~\bibnamefont{Onogi}}, \bibnamefont{and}
  \bibinfo{author}{\bibfnamefont{N.}~\bibnamefont{Yamada}},
  \bibinfo{journal}{PoS} \textbf{\bibinfo{volume}{LAT2006}},
  \bibinfo{pages}{031} (\bibinfo{year}{2006}), \eprint{hep-lat/0610026}.

\bibitem[{\citenamefont{Aoki et~al.}(2008)\citenamefont{Aoki, Fukaya,
  Hashimoto, Ishikawa, Kanaya, Kaneko, Matsufuru, Okamoto, Okawa, Onogi
  et~al.}}]{AoFuHa08}
\bibinfo{author}{\bibfnamefont{S.}~\bibnamefont{Aoki}},
  \bibinfo{author}{\bibfnamefont{H.}~\bibnamefont{Fukaya}},
  \bibinfo{author}{\bibfnamefont{S.}~\bibnamefont{Hashimoto}},
  \bibinfo{author}{\bibfnamefont{K.-I.} \bibnamefont{Ishikawa}},
  \bibinfo{author}{\bibfnamefont{K.}~\bibnamefont{Kanaya}},
  \bibinfo{author}{\bibfnamefont{T.}~\bibnamefont{Kaneko}},
  \bibinfo{author}{\bibfnamefont{H.}~\bibnamefont{Matsufuru}},
  \bibinfo{author}{\bibfnamefont{M.}~\bibnamefont{Okamoto}},
  \bibinfo{author}{\bibfnamefont{M.}~\bibnamefont{Okawa}},
  \bibinfo{author}{\bibfnamefont{T.}~\bibnamefont{Onogi}}, \bibnamefont{et~al.}
  (\bibinfo{year}{2008}), \eprint{arXiv:0803.3197 [hep-lat]}.

\bibitem[{\citenamefont{Noaki et~al.}(2007)\citenamefont{Noaki, Aoki, Fukaya,
  Hashimoto, Kaneko, Matsufuru, Onogi, Shintani, and Yamada}}]{NoAoFu07}
\bibinfo{author}{\bibfnamefont{J.}~\bibnamefont{Noaki}},
  \bibinfo{author}{\bibfnamefont{S.}~\bibnamefont{Aoki}},
  \bibinfo{author}{\bibfnamefont{H.}~\bibnamefont{Fukaya}},
  \bibinfo{author}{\bibfnamefont{S.}~\bibnamefont{Hashimoto}},
  \bibinfo{author}{\bibfnamefont{T.}~\bibnamefont{Kaneko}},
  \bibinfo{author}{\bibfnamefont{H.}~\bibnamefont{Matsufuru}},
  \bibinfo{author}{\bibfnamefont{T.}~\bibnamefont{Onogi}},
  \bibinfo{author}{\bibfnamefont{E.}~\bibnamefont{Shintani}}, \bibnamefont{and}
  \bibinfo{author}{\bibfnamefont{N.}~\bibnamefont{Yamada}},
  \bibinfo{journal}{PoS} \textbf{\bibinfo{volume}{LATTICE 2007}},
  \bibinfo{pages}{126} (\bibinfo{year}{2007}),
\eprint{arXiv:0710.0929 [hep-lat]}.

\bibitem[{\citenamefont{Cundy et~al.}(2005{\natexlab{a}})\citenamefont{Cundy,
  Krieg, and Lippert}}]{CuKrLi05}
\bibinfo{author}{\bibfnamefont{N.}~\bibnamefont{Cundy}},
  \bibinfo{author}{\bibfnamefont{S.}~\bibnamefont{Krieg}}, \bibnamefont{and}
  \bibinfo{author}{\bibfnamefont{T.}~\bibnamefont{Lippert}},
  \bibinfo{journal}{PoS} \textbf{\bibinfo{volume}{LAT2005}},
  \bibinfo{pages}{107} (\bibinfo{year}{2005}{\natexlab{a}}),
\eprint{hep-lat/0511044}.

\bibitem[{\citenamefont{Cundy et~al.}(2005{\natexlab{b}})\citenamefont{Cundy,
  Krieg, Arnold, Frommer, Lippert, and Schilling}}]{CuKrAr05}
\bibinfo{author}{\bibfnamefont{N.}~\bibnamefont{Cundy}},
  \bibinfo{author}{\bibfnamefont{S.}~\bibnamefont{Krieg}},
  \bibinfo{author}{\bibfnamefont{G.}~\bibnamefont{Arnold}},
  \bibinfo{author}{\bibfnamefont{A.}~\bibnamefont{Frommer}},
  \bibinfo{author}{\bibfnamefont{T.}~\bibnamefont{Lippert}}, \bibnamefont{and}
  \bibinfo{author}{\bibfnamefont{K.}~\bibnamefont{Schilling}}
  (\bibinfo{year}{2005}{\natexlab{b}}), \eprint{hep-lat/0502007}.

\bibitem[{\citenamefont{Cundy et~al.}(2007)\citenamefont{Cundy, Krieg, Lippert,
  and Sch{\"a}fer}}]{CuKrLi07}
\bibinfo{author}{\bibfnamefont{N.}~\bibnamefont{Cundy}},
  \bibinfo{author}{\bibfnamefont{S.}~\bibnamefont{Krieg}},
  \bibinfo{author}{\bibfnamefont{T.}~\bibnamefont{Lippert}}, \bibnamefont{and}
  \bibinfo{author}{\bibfnamefont{A.}~\bibnamefont{Sch{\"a}fer}},
  \bibinfo{journal}{PoS} \textbf{\bibinfo{volume}{LATTICE2007}},
  \bibinfo{pages}{030} (\bibinfo{year}{2007}),
\eprint{arXiv:0710.1785 [hep-lat]}.

\bibitem[{\citenamefont{Cundy et~al.}(2008)\citenamefont{Cundy, Krieg, Lippert,
  and Sch{\"a}fer}}]{CuKrLi08}
\bibinfo{author}{\bibfnamefont{N.}~\bibnamefont{Cundy}},
  \bibinfo{author}{\bibfnamefont{S.}~\bibnamefont{Krieg}},
  \bibinfo{author}{\bibfnamefont{T.}~\bibnamefont{Lippert}}, \bibnamefont{and}
  \bibinfo{author}{\bibfnamefont{A.}~\bibnamefont{Sch{\"a}fer}}
  (\bibinfo{year}{2008}), \eprint{arXiv:0803.0294 [hep-lat]}.

\bibitem[{\citenamefont{DeGrand and Schaefer}(2005{\natexlab{a}})}]{DeSc05a}
\bibinfo{author}{\bibfnamefont{T.}~\bibnamefont{DeGrand}} \bibnamefont{and}
  \bibinfo{author}{\bibfnamefont{S.}~\bibnamefont{Schaefer}},
  \bibinfo{journal}{Phys. Rev. D} \textbf{\bibinfo{volume}{71}},
  \bibinfo{pages}{034507} (\bibinfo{year}{2005}{\natexlab{a}}),
\eprint{hep-lat/0412005}.

\bibitem[{\citenamefont{DeGrand and Schaefer}(2005{\natexlab{b}})}]{DeSc05b}
\bibinfo{author}{\bibfnamefont{T.}~\bibnamefont{DeGrand}} \bibnamefont{and}
  \bibinfo{author}{\bibfnamefont{S.}~\bibnamefont{Schaefer}},
  \bibinfo{journal}{Phys. Rev. D} \textbf{\bibinfo{volume}{72}},
  \bibinfo{pages}{054503} (\bibinfo{year}{2005}{\natexlab{b}}),
  \eprint{hep-lat/0506021}.

\bibitem[{\citenamefont{Fodor et~al.}(2005)\citenamefont{Fodor, Katz, and
  Szabo}}]{FoKaSz05}
\bibinfo{author}{\bibfnamefont{Z.}~\bibnamefont{Fodor}},
  \bibinfo{author}{\bibfnamefont{S.~D.} \bibnamefont{Katz}}, \bibnamefont{and}
  \bibinfo{author}{\bibfnamefont{K.~K.} \bibnamefont{Szabo}},
  \bibinfo{journal}{Nucl. Phys. B (Proc. Suppl.)}
  \textbf{\bibinfo{volume}{140}}, \bibinfo{pages}{704} (\bibinfo{year}{2005}),
  \eprint{hep-lat/0409070}.

\bibitem[{\citenamefont{Egri et~al.}(2006)\citenamefont{Egri, Fodor, Katz, and
  Szabo}}]{EgFoKa05}
\bibinfo{author}{\bibfnamefont{G.~I.} \bibnamefont{Egri}},
  \bibinfo{author}{\bibfnamefont{Z.}~\bibnamefont{Fodor}},
  \bibinfo{author}{\bibfnamefont{S.~D.} \bibnamefont{Katz}}, \bibnamefont{and}
  \bibinfo{author}{\bibfnamefont{K.}~\bibnamefont{Szabo}},
  \bibinfo{journal}{JHEP} \textbf{\bibinfo{volume}{0601}}, \bibinfo{pages}{049}
  (\bibinfo{year}{2006}),
\eprint{hep-lat/0510117}.

\bibitem[{\citenamefont{Gattringer et~al.}(2004)\citenamefont{Gattringer,
  G{\"o}ckeler, Hasenfratz, Hauswirth, Holland, J{\"o}rg, Juge, Lang,
  Niedermayer, Rakow et~al.}}]{GaGoHa03a}
\bibinfo{author}{\bibfnamefont{C.}~\bibnamefont{Gattringer}},
  \bibinfo{author}{\bibfnamefont{M.}~\bibnamefont{G{\"o}ckeler}},
  \bibinfo{author}{\bibfnamefont{P.}~\bibnamefont{Hasenfratz}},
  \bibinfo{author}{\bibfnamefont{S.}~\bibnamefont{Hauswirth}},
  \bibinfo{author}{\bibfnamefont{K.}~\bibnamefont{Holland}},
  \bibinfo{author}{\bibfnamefont{T.}~\bibnamefont{J{\"o}rg}},
  \bibinfo{author}{\bibfnamefont{K.~J.} \bibnamefont{Juge}},
  \bibinfo{author}{\bibfnamefont{C.~B.} \bibnamefont{Lang}},
  \bibinfo{author}{\bibfnamefont{F.}~\bibnamefont{Niedermayer}},
  \bibinfo{author}{\bibfnamefont{P.~E.~L.} \bibnamefont{Rakow}},
  \bibnamefont{et~al.}, \bibinfo{journal}{Nucl. Phys.}
  \textbf{\bibinfo{volume}{B677}}, \bibinfo{pages}{3} (\bibinfo{year}{2004}),
  \eprint{hep-lat/0307013}.

\bibitem[{\citenamefont{Gattringer et~al.}(2005)\citenamefont{Gattringer,
  Huber, and Lang}}]{GaHuLa05a}
\bibinfo{author}{\bibfnamefont{C.}~\bibnamefont{Gattringer}},
  \bibinfo{author}{\bibfnamefont{P.}~\bibnamefont{Huber}}, \bibnamefont{and}
  \bibinfo{author}{\bibfnamefont{C.~B.} \bibnamefont{Lang}},
  \bibinfo{journal}{Phys. Rev. D} \textbf{\bibinfo{volume}{72}},
  \bibinfo{pages}{094510} (\bibinfo{year}{2005}), \eprint{hep-lat/0509003}.

\bibitem[{\citenamefont{Lang et~al.}(2006)\citenamefont{Lang, Majumdar, and
  Ortner}}]{LaMaOr05c}
\bibinfo{author}{\bibfnamefont{C.~B.} \bibnamefont{Lang}},
  \bibinfo{author}{\bibfnamefont{P.}~\bibnamefont{Majumdar}}, \bibnamefont{and}
  \bibinfo{author}{\bibfnamefont{W.}~\bibnamefont{Ortner}},
  \bibinfo{journal}{Phys. Rev. D} \textbf{\bibinfo{volume}{73}},
  \bibinfo{pages}{034507} (\bibinfo{year}{2006}), \eprint{hep-lat/0512014}.

\bibitem[{\citenamefont{Frigori et~al.}(2007)\citenamefont{Frigori, Gattringer,
  Lang, Limmer, Maurer, Mohler, and Sch{\"a}fer}}]{FrGaLa07}
\bibinfo{author}{\bibfnamefont{R.}~\bibnamefont{Frigori}},
  \bibinfo{author}{\bibfnamefont{C.}~\bibnamefont{Gattringer}},
  \bibinfo{author}{\bibfnamefont{C.~B.} \bibnamefont{Lang}},
  \bibinfo{author}{\bibfnamefont{M.}~\bibnamefont{Limmer}},
  \bibinfo{author}{\bibfnamefont{T.}~\bibnamefont{Maurer}},
  \bibinfo{author}{\bibfnamefont{D.}~\bibnamefont{Mohler}}, \bibnamefont{and}
  \bibinfo{author}{\bibfnamefont{A.}~\bibnamefont{Sch{\"a}fer}},
  \bibinfo{journal}{PoS} \textbf{\bibinfo{volume}{LATTICE2007}},
  \bibinfo{pages}{114} (\bibinfo{year}{2007}), \eprint{arXiv:0709.4582v1
  [hep-lat]}.

\bibitem[{\citenamefont{Lang}(2008)}]{La08}
\bibinfo{author}{\bibfnamefont{C.~B.} \bibnamefont{Lang}},
  \bibinfo{journal}{Prog. Part. Nucl. Phys.} \textbf{\bibinfo{volume}{61}},
  \bibinfo{pages}{35} (\bibinfo{year}{2008}), \eprint{arXiv:0711.3091
  [nucl-th]}.

\bibitem[{\citenamefont{Lang et~al.}(2005)\citenamefont{Lang, Majumdar, and
  Ortner}}]{LaMaOr05b}
\bibinfo{author}{\bibfnamefont{C.~B.} \bibnamefont{Lang}},
  \bibinfo{author}{\bibfnamefont{P.}~\bibnamefont{Majumdar}}, \bibnamefont{and}
  \bibinfo{author}{\bibfnamefont{W.}~\bibnamefont{Ortner}},
  \bibinfo{journal}{PoS} \textbf{\bibinfo{volume}{LAT2005}},
  \bibinfo{pages}{131} (\bibinfo{year}{2005}), \eprint{hep-lat/0509005}.

\bibitem[{\citenamefont{L{\"u}scher and Weisz}(1985)}]{LuWe85}
\bibinfo{author}{\bibfnamefont{M.}~\bibnamefont{L{\"u}scher}} \bibnamefont{and}
  \bibinfo{author}{\bibfnamefont{P.}~\bibnamefont{Weisz}},
  \bibinfo{journal}{Commun. Math. Phys.} \textbf{\bibinfo{volume}{97}},
  \bibinfo{pages}{59}
 (\bibinfo{year}{1985}).

\bibitem[{\citenamefont{D{\"u}rr and Hoelbling}(2004)}]{DuHo04}
\bibinfo{author}{\bibfnamefont{S.}~\bibnamefont{D{\"u}rr}} \bibnamefont{and}
  \bibinfo{author}{\bibfnamefont{C.}~\bibnamefont{Hoelbling}},
  \bibinfo{journal}{Phys. Rev. D} \textbf{\bibinfo{volume}{69}},
  \bibinfo{pages}{034503} (\bibinfo{year}{2004}), \eprint{hep-lat/0311002}.

\bibitem[{\citenamefont{Hasenfratz and Knechtli}(2001)}]{HaKn01}
\bibinfo{author}{\bibfnamefont{A.}~\bibnamefont{Hasenfratz}} \bibnamefont{and}
  \bibinfo{author}{\bibfnamefont{F.}~\bibnamefont{Knechtli}},
  \bibinfo{journal}{Phys. Rev. D} \textbf{\bibinfo{volume}{64}},
  \bibinfo{pages}{034504}
 (\bibinfo{year}{2001}).

\bibitem[{\citenamefont{Morningstar and Peardon}(2004)}]{MoPe04}
\bibinfo{author}{\bibfnamefont{C.}~\bibnamefont{Morningstar}} \bibnamefont{and}
  \bibinfo{author}{\bibfnamefont{M.}~\bibnamefont{Peardon}},
  \bibinfo{journal}{Phys. Rev. D} \textbf{\bibinfo{volume}{69}},
  \bibinfo{pages}{054501} (\bibinfo{year}{2004}),
\eprint{hep-lat/0311018}.

\bibitem[{\citenamefont{Hasenfratz et~al.}(2007)\citenamefont{Hasenfratz,
  Hoffmann, and Schaefer}}]{HaHoSc07}
\bibinfo{author}{\bibfnamefont{A.}~\bibnamefont{Hasenfratz}},
  \bibinfo{author}{\bibfnamefont{R.}~\bibnamefont{Hoffmann}}, \bibnamefont{and}
  \bibinfo{author}{\bibfnamefont{S.}~\bibnamefont{Schaefer}},
  \bibinfo{journal}{JHEP} \textbf{\bibinfo{volume}{0705}}, \bibinfo{pages}{029}
 (\bibinfo{year}{2007}).

\bibitem[{\citenamefont{D{\"u}rr}(2007)}]{Du07}
\bibinfo{author}{\bibfnamefont{S.}~\bibnamefont{D{\"u}rr}}
  (\bibinfo{year}{2007}), \eprint{arXiv:0709.4110 [hep-lat]}.

\bibitem[{\citenamefont{Duane et~al.}(1987)\citenamefont{Duane, Kennedy,
  Pendleton, and Roweth}}]{DuKePe87}
\bibinfo{author}{\bibfnamefont{S.}~\bibnamefont{Duane}},
  \bibinfo{author}{\bibfnamefont{A.~D.} \bibnamefont{Kennedy}},
  \bibinfo{author}{\bibfnamefont{B.~J.} \bibnamefont{Pendleton}},
  \bibnamefont{and} \bibinfo{author}{\bibfnamefont{D.}~\bibnamefont{Roweth}},
  \bibinfo{journal}{Phys. Lett. B} \textbf{\bibinfo{volume}{195}},
  \bibinfo{pages}{216} (\bibinfo{year}{1987}).

\bibitem[{\citenamefont{Ortner}(2006)}]{Or06}
\bibinfo{author}{\bibfnamefont{W.}~\bibnamefont{Ortner}}, Ph.D. thesis,
  \bibinfo{school}{Univ. Graz} (\bibinfo{year}{2006}).

\bibitem[{\citenamefont{Hasenbusch}(2001)}]{Ha01a}
\bibinfo{author}{\bibfnamefont{M.}~\bibnamefont{Hasenbusch}},
  \bibinfo{journal}{Phys. Lett. B} \textbf{\bibinfo{volume}{519}},
  \bibinfo{pages}{177} (\bibinfo{year}{2001}),
\eprint{hep-lat/0107019}.

\bibitem[{\citenamefont{Brower et~al.}(1997)\citenamefont{Brower, Ivanenko,
  Levi, and Orginos}}]{BrIvLe97}
\bibinfo{author}{\bibfnamefont{R.~C.} \bibnamefont{Brower}},
  \bibinfo{author}{\bibfnamefont{T.}~\bibnamefont{Ivanenko}},
  \bibinfo{author}{\bibfnamefont{A.~R.} \bibnamefont{Levi}}, \bibnamefont{and}
  \bibinfo{author}{\bibfnamefont{K.~N.} \bibnamefont{Orginos}},
  \bibinfo{journal}{Nucl. Phys.} \textbf{\bibinfo{volume}{B484}},
  \bibinfo{pages}{353} (\bibinfo{year}{1997}), \eprint{hep-lat/9509012}.

\bibitem[{\citenamefont{D\"urr et~al.}(2008)\citenamefont{D\"urr, Fodor,
  Hoelbling, Hoffmann, Katz, Krieg, Kurth, Lellouch, Lippert, Szabo
  et~al.}}]{DuFoHo08}
\bibinfo{author}{\bibfnamefont{S.}~\bibnamefont{D\"urr}},
  \bibinfo{author}{\bibfnamefont{Z.}~\bibnamefont{Fodor}},
  \bibinfo{author}{\bibfnamefont{C.}~\bibnamefont{Hoelbling}},
  \bibinfo{author}{\bibfnamefont{R.}~\bibnamefont{Hoffmann}},
  \bibinfo{author}{\bibfnamefont{S.}~\bibnamefont{Katz}},
  \bibinfo{author}{\bibfnamefont{S.}~\bibnamefont{Krieg}},
  \bibinfo{author}{\bibfnamefont{T.}~\bibnamefont{Kurth}},
  \bibinfo{author}{\bibfnamefont{L.}~\bibnamefont{Lellouch}},
  \bibinfo{author}{\bibfnamefont{T.}~\bibnamefont{Lippert}},
  \bibinfo{author}{\bibfnamefont{K.}~\bibnamefont{Szabo}}, \bibnamefont{et~al.}
  (\bibinfo{year}{2008}), \eprint{arXiv:0802.2706 [hep-lat]}.

\bibitem[{\citenamefont{Creutz}(1988)}]{Cr88}
\bibinfo{author}{\bibfnamefont{M.}~\bibnamefont{Creutz}},
  \bibinfo{journal}{Phys. Rev. D} \textbf{\bibinfo{volume}{38}},
  \bibinfo{pages}{1228}
 (\bibinfo{year}{1988}).

\bibitem[{\citenamefont{Namekawa et~al.}(2004)\citenamefont{Namekawa, Aoki,
  Fukugita, Ishikawa, Ishizuka, Iwasaki, Kanaya, Kaneko, Kuramashi, Lesk
  et~al.}}]{NaAoFu04}
\bibinfo{author}{\bibfnamefont{Y.}~\bibnamefont{Namekawa}},
  \bibinfo{author}{\bibfnamefont{S.}~\bibnamefont{Aoki}},
  \bibinfo{author}{\bibfnamefont{M.}~\bibnamefont{Fukugita}},
  \bibinfo{author}{\bibfnamefont{K.-I.} \bibnamefont{Ishikawa}},
  \bibinfo{author}{\bibfnamefont{N.}~\bibnamefont{Ishizuka}},
  \bibinfo{author}{\bibfnamefont{Y.}~\bibnamefont{Iwasaki}},
  \bibinfo{author}{\bibfnamefont{K.}~\bibnamefont{Kanaya}},
  \bibinfo{author}{\bibfnamefont{T.}~\bibnamefont{Kaneko}},
  \bibinfo{author}{\bibfnamefont{Y.}~\bibnamefont{Kuramashi}},
  \bibinfo{author}{\bibfnamefont{V.~I.} \bibnamefont{Lesk}},
  \bibnamefont{et~al.}, \bibinfo{journal}{Phys. Rev. D}
  \textbf{\bibinfo{volume}{70}}, \bibinfo{pages}{074503}
  (\bibinfo{year}{2004}), \eprint{hep-lat/0404014}.

\bibitem[{\citenamefont{Joo et~al.}(2000)\citenamefont{Joo, Pendleton, Kennedy,
  Irving, Sexton, Pickles, and Booth}}]{JoPeKe00}
\bibinfo{author}{\bibfnamefont{B.}~\bibnamefont{Joo}},
  \bibinfo{author}{\bibfnamefont{B.}~\bibnamefont{Pendleton}},
  \bibinfo{author}{\bibfnamefont{A.~D.} \bibnamefont{Kennedy}},
  \bibinfo{author}{\bibfnamefont{A.~C.} \bibnamefont{Irving}},
  \bibinfo{author}{\bibfnamefont{J.~C.} \bibnamefont{Sexton}},
  \bibinfo{author}{\bibfnamefont{S.~M.} \bibnamefont{Pickles}},
  \bibnamefont{and} \bibinfo{author}{\bibfnamefont{S.~P.} \bibnamefont{Booth}},
  \bibinfo{journal}{Phys. Rev. D} \textbf{\bibinfo{volume}{62}},
  \bibinfo{pages}{114501} (\bibinfo{year}{2000}), \eprint{hep-lat/0005023}.

\bibitem[{\citenamefont{Gupta et~al.}(1990)\citenamefont{Gupta, Irback, Karsch,
  and Petersson}}]{GuIrKa90}
\bibinfo{author}{\bibfnamefont{S.}~\bibnamefont{Gupta}},
  \bibinfo{author}{\bibfnamefont{A.}~\bibnamefont{Irback}},
  \bibinfo{author}{\bibfnamefont{F.}~\bibnamefont{Karsch}}, \bibnamefont{and}
  \bibinfo{author}{\bibfnamefont{B.}~\bibnamefont{Petersson}},
  \bibinfo{journal}{Phys. Lett. B} \textbf{\bibinfo{volume}{242}},
  \bibinfo{pages}{437}
 (\bibinfo{year}{1990}).

\bibitem[{\citenamefont{Atiyah and Singer}(1971)}]{AtSi71}
\bibinfo{author}{\bibfnamefont{M.}~\bibnamefont{Atiyah}} \bibnamefont{and}
  \bibinfo{author}{\bibfnamefont{I.~M.} \bibnamefont{Singer}},
  \bibinfo{journal}{Ann. Math.} \textbf{\bibinfo{volume}{93}},
  \bibinfo{pages}{139}
 (\bibinfo{year}{1971}).

\bibitem[{\citenamefont{Sommer}(1994)}]{So94}
\bibinfo{author}{\bibfnamefont{R.}~\bibnamefont{Sommer}},
  \bibinfo{journal}{Nucl. Phys.} \textbf{\bibinfo{volume}{B411}},
  \bibinfo{pages}{839} (\bibinfo{year}{1994}), \eprint{hep-lat/9310022}.

\bibitem[{\citenamefont{Lang and Rebbi}(1982)}]{LaRe82}
\bibinfo{author}{\bibfnamefont{C.~B.} \bibnamefont{Lang}} \bibnamefont{and}
  \bibinfo{author}{\bibfnamefont{C.}~\bibnamefont{Rebbi}},
  \bibinfo{journal}{Phys. Lett.} \textbf{\bibinfo{volume}{115B}},
  \bibinfo{pages}{137}
 (\bibinfo{year}{1982}).

\bibitem[{\citenamefont{Michael}(1985)}]{Mi85}
\bibinfo{author}{\bibfnamefont{C.}~\bibnamefont{Michael}},
  \bibinfo{journal}{Nucl. Phys.} \textbf{\bibinfo{volume}{B259}},
  \bibinfo{pages}{58}
 (\bibinfo{year}{1985}).

\bibitem[{\citenamefont{Edwards et~al.}(1998)\citenamefont{Edwards, Heller, and
  Klassen}}]{EdHeKl98}
\bibinfo{author}{\bibfnamefont{R.~G.} \bibnamefont{Edwards}},
  \bibinfo{author}{\bibfnamefont{U.~M.} \bibnamefont{Heller}},
  \bibnamefont{and} \bibinfo{author}{\bibfnamefont{T.~R.}
  \bibnamefont{Klassen}}, \bibinfo{journal}{Nucl. Phys.}
  \textbf{\bibinfo{volume}{B517}}, \bibinfo{pages}{377} (\bibinfo{year}{1998}),
\eprint{hep-lat/9711003}.

\bibitem[{\citenamefont{Allton et~al.}(2002)\citenamefont{Allton, Booth,
  Bowler, Garden, Hart, Hepburn, Irving, Jo\'o, Kenway, Maynard
  et~al.}}]{AlBoBo02}
\bibinfo{author}{\bibfnamefont{C.~R.} \bibnamefont{Allton}},
  \bibinfo{author}{\bibfnamefont{S.~P.} \bibnamefont{Booth}},
  \bibinfo{author}{\bibfnamefont{K.~C.} \bibnamefont{Bowler}},
  \bibinfo{author}{\bibfnamefont{J.}~\bibnamefont{Garden}},
  \bibinfo{author}{\bibfnamefont{A.}~\bibnamefont{Hart}},
  \bibinfo{author}{\bibfnamefont{D.}~\bibnamefont{Hepburn}},
  \bibinfo{author}{\bibfnamefont{A.~C.} \bibnamefont{Irving}},
  \bibinfo{author}{\bibfnamefont{B.}~\bibnamefont{Jo\'o}},
  \bibinfo{author}{\bibfnamefont{R.~D.} \bibnamefont{Kenway}},
  \bibinfo{author}{\bibfnamefont{C.~M.} \bibnamefont{Maynard}},
  \bibnamefont{et~al.}, \bibinfo{journal}{Phys. Rev. D}
  \textbf{\bibinfo{volume}{65}}, \bibinfo{pages}{054502}
  (\bibinfo{year}{2002}),
\eprint{hep-lat/0107021}.

\bibitem[{\citenamefont{Hasenfratz et~al.}(2002)\citenamefont{Hasenfratz,
  Hoffmann, and Knechtli}}]{HaHoKn01}
\bibinfo{author}{\bibfnamefont{A.}~\bibnamefont{Hasenfratz}},
  \bibinfo{author}{\bibfnamefont{R.}~\bibnamefont{Hoffmann}}, \bibnamefont{and}
  \bibinfo{author}{\bibfnamefont{F.}~\bibnamefont{Knechtli}},
  \bibinfo{journal}{Nucl. Phys. B (Proc. Suppl.)}
  \textbf{\bibinfo{volume}{106}}, \bibinfo{pages}{418} (\bibinfo{year}{2002}),
  \eprint{hep-lat/0110168}.

\bibitem[{\citenamefont{Gattringer et~al.}(2002)\citenamefont{Gattringer,
  Hoffmann, and Schaefer}}]{GaHoSc02a}
\bibinfo{author}{\bibfnamefont{C.}~\bibnamefont{Gattringer}},
  \bibinfo{author}{\bibfnamefont{R.}~\bibnamefont{Hoffmann}}, \bibnamefont{and}
  \bibinfo{author}{\bibfnamefont{S.}~\bibnamefont{Schaefer}},
  \bibinfo{journal}{Phys. Rev. D} \textbf{\bibinfo{volume}{65}},
  \bibinfo{pages}{094503} (\bibinfo{year}{2002}), \eprint{hep-lat/0112024}.

\bibitem[{\citenamefont{Aoki}(2001)}]{Ao01}
\bibinfo{author}{\bibfnamefont{S.}~\bibnamefont{Aoki}}, \bibinfo{journal}{Nucl.
  Phys. B (Proc. Suppl.)} \textbf{\bibinfo{volume}{94}}, \bibinfo{pages}{3}
  (\bibinfo{year}{2001}), \eprint{hep-lat/0011074}.

\bibitem[{\citenamefont{Michael}(1994)}]{Mi94}
\bibinfo{author}{\bibfnamefont{C.}~\bibnamefont{Michael}},
  \bibinfo{journal}{Phys. Rev. D} \textbf{\bibinfo{volume}{49}},
  \bibinfo{pages}{2616} (\bibinfo{year}{1994}), \eprint{hep-lat/9310026}.

\bibitem[{\citenamefont{Michael and McKerrell}(1995)}]{McMi95}
\bibinfo{author}{\bibfnamefont{C.}~\bibnamefont{Michael}} \bibnamefont{and}
  \bibinfo{author}{\bibfnamefont{A.}~\bibnamefont{McKerrell}},
  \bibinfo{journal}{Phys. Rev. D} \textbf{\bibinfo{volume}{51}},
  \bibinfo{pages}{3745} (\bibinfo{year}{1995}), \eprint{hep-lat/9412087}.

\bibitem[{\citenamefont{Asakawa et~al.}(2001)\citenamefont{Asakawa, Hatsuda,
  and Nakahara}}]{AsHaNa01}
\bibinfo{author}{\bibfnamefont{M.}~\bibnamefont{Asakawa}},
  \bibinfo{author}{\bibfnamefont{T.}~\bibnamefont{Hatsuda}}, \bibnamefont{and}
  \bibinfo{author}{\bibfnamefont{Y.}~\bibnamefont{Nakahara}},
  \bibinfo{journal}{Prog. Part. Nucl. Phys.} \textbf{\bibinfo{volume}{46}},
  \bibinfo{pages}{459} (\bibinfo{year}{2001}), \eprint{hep-lat/0011040}.

\bibitem[{\citenamefont{Lepage et~al.}(2002)\citenamefont{Lepage, Clark,
  Davies, Hornbostel, Mackenzie, Morningstar, and Trottier}}]{LeClDa01}
\bibinfo{author}{\bibfnamefont{G.~P.} \bibnamefont{Lepage}},
  \bibinfo{author}{\bibfnamefont{B.}~\bibnamefont{Clark}},
  \bibinfo{author}{\bibfnamefont{C.~T.~H.} \bibnamefont{Davies}},
  \bibinfo{author}{\bibfnamefont{K.}~\bibnamefont{Hornbostel}},
  \bibinfo{author}{\bibfnamefont{P.~B.} \bibnamefont{Mackenzie}},
  \bibinfo{author}{\bibfnamefont{C.}~\bibnamefont{Morningstar}},
  \bibnamefont{and} \bibinfo{author}{\bibfnamefont{H.}~\bibnamefont{Trottier}},
  \bibinfo{journal}{Nucl. Phys. Proc. Suppl.} \textbf{\bibinfo{volume}{106}},
  \bibinfo{pages}{12} (\bibinfo{year}{2002}), \eprint{hep-lat/0110175}.

\bibitem[{\citenamefont{Chen et~al.}(2004)\citenamefont{Chen, Dong, Draper,
  Horv{\'a}th, Liu, Mathur, Tamhankar, Srinivasan, Lee, and Zhang}}]{ChDoDr04}
\bibinfo{author}{\bibfnamefont{Y.}~\bibnamefont{Chen}},
  \bibinfo{author}{\bibfnamefont{S.-J.} \bibnamefont{Dong}},
  \bibinfo{author}{\bibfnamefont{T.}~\bibnamefont{Draper}},
  \bibinfo{author}{\bibfnamefont{I.}~\bibnamefont{Horv{\'a}th}},
  \bibinfo{author}{\bibfnamefont{K.-F.} \bibnamefont{Liu}},
  \bibinfo{author}{\bibfnamefont{N.}~\bibnamefont{Mathur}},
  \bibinfo{author}{\bibfnamefont{S.}~\bibnamefont{Tamhankar}},
  \bibinfo{author}{\bibfnamefont{C.}~\bibnamefont{Srinivasan}},
  \bibinfo{author}{\bibfnamefont{F.~X.} \bibnamefont{Lee}}, \bibnamefont{and}
  \bibinfo{author}{\bibfnamefont{J.}~\bibnamefont{Zhang}}
  (\bibinfo{year}{2004}), \eprint{hep-lat/0405001}.

\bibitem[{\citenamefont{Guadagnoli et~al.}(2004)\citenamefont{Guadagnoli,
  Papinutto, and Simula}}]{GuPaSi04}
\bibinfo{author}{\bibfnamefont{D.}~\bibnamefont{Guadagnoli}},
  \bibinfo{author}{\bibfnamefont{M.}~\bibnamefont{Papinutto}},
  \bibnamefont{and} \bibinfo{author}{\bibfnamefont{S.}~\bibnamefont{Simula}},
  \bibinfo{journal}{Phys. Lett. B} \textbf{\bibinfo{volume}{604}},
  \bibinfo{pages}{74} (\bibinfo{year}{2004}),
\eprint{hep-lat/0409011}.

\bibitem[{\citenamefont{von Hippel et~al.}(2008)\citenamefont{von Hippel,
  Lewis, and Petry}}]{HiLePe07a}
\bibinfo{author}{\bibfnamefont{G.~M.} \bibnamefont{von Hippel}},
  \bibinfo{author}{\bibfnamefont{R.}~\bibnamefont{Lewis}}, \bibnamefont{and}
  \bibinfo{author}{\bibfnamefont{R.~G.} \bibnamefont{Petry}},
  \bibinfo{journal}{Comput. Phys. Commun.} \textbf{\bibinfo{volume}{178}},
  \bibinfo{pages}{713} (\bibinfo{year}{2008}), \eprint{arXiv:0707.2788
  [hep-lat]}.

\bibitem[{\citenamefont{von Hippel et~al.}(2007)\citenamefont{von Hippel,
  Lewis, and Petry}}]{HiLePe07}
\bibinfo{author}{\bibfnamefont{G.~M.} \bibnamefont{von Hippel}},
  \bibinfo{author}{\bibfnamefont{R.}~\bibnamefont{Lewis}}, \bibnamefont{and}
  \bibinfo{author}{\bibfnamefont{R.~G.} \bibnamefont{Petry}},
  \bibinfo{journal}{PoS} \textbf{\bibinfo{volume}{LATTICE2007}},
  \bibinfo{pages}{043} (\bibinfo{year}{2007}), \eprint{arXiv:0710.0014
  [hep-lat]}.

\bibitem[{\citenamefont{L{\"u}scher and Wolff}(1990)}]{LuWo90}
\bibinfo{author}{\bibfnamefont{M.}~\bibnamefont{L{\"u}scher}} \bibnamefont{and}
  \bibinfo{author}{\bibfnamefont{U.}~\bibnamefont{Wolff}},
  \bibinfo{journal}{Nucl. Pbys.} \textbf{\bibinfo{volume}{B339}},
  \bibinfo{pages}{222} (\bibinfo{year}{1990}).

\bibitem[{\citenamefont{Burch et~al.}(2004)\citenamefont{Burch, Gattringer,
  Glozman, Kleindl, Lang, and Sch{\"a}fer}}]{BuGaGl04a}
\bibinfo{author}{\bibfnamefont{T.}~\bibnamefont{Burch}},
  \bibinfo{author}{\bibfnamefont{C.}~\bibnamefont{Gattringer}},
  \bibinfo{author}{\bibfnamefont{L.~Y.} \bibnamefont{Glozman}},
  \bibinfo{author}{\bibfnamefont{R.}~\bibnamefont{Kleindl}},
  \bibinfo{author}{\bibfnamefont{C.~B.} \bibnamefont{Lang}}, \bibnamefont{and}
  \bibinfo{author}{\bibfnamefont{A.}~\bibnamefont{Sch{\"a}fer}},
  \bibinfo{journal}{Phys. Rev. D} \textbf{\bibinfo{volume}{70}},
  \bibinfo{pages}{054502} (\bibinfo{year}{2004}), \eprint{hep-lat/0405006}.

\bibitem[{\citenamefont{Burch et~al.}(2005{\natexlab{a}})\citenamefont{Burch,
  Gattringer, Glozman, Kleindl, Lang, , and Sch{\"a}fer}}]{BuGaGl05a}
\bibinfo{author}{\bibfnamefont{T.}~\bibnamefont{Burch}},
  \bibinfo{author}{\bibfnamefont{C.}~\bibnamefont{Gattringer}},
  \bibinfo{author}{\bibfnamefont{L.~Y.} \bibnamefont{Glozman}},
  \bibinfo{author}{\bibfnamefont{R.}~\bibnamefont{Kleindl}},
  \bibinfo{author}{\bibfnamefont{C.~B.} \bibnamefont{Lang}},
  \bibnamefont{and}
  \bibinfo{author}{\bibfnamefont{A.}~\bibnamefont{Sch{\"a}fer}},
  \bibinfo{journal}{Nucl. Phys. B (Proc. Suppl.)}
  \textbf{\bibinfo{volume}{140}}, \bibinfo{pages}{284}
  (\bibinfo{year}{2005}{\natexlab{a}}), \eprint{hep-lat/0409014}.

\bibitem[{\citenamefont{Burch et~al.}(2005{\natexlab{b}})\citenamefont{Burch,
  Gattringer, Glozman, Hagen, Hierl, Lang, and Sch{\"a}fer}}]{BuGaGl05b}
\bibinfo{author}{\bibfnamefont{T.}~\bibnamefont{Burch}},
  \bibinfo{author}{\bibfnamefont{C.}~\bibnamefont{Gattringer}},
  \bibinfo{author}{\bibfnamefont{L.~Y.} \bibnamefont{Glozman}},
  \bibinfo{author}{\bibfnamefont{C.}~\bibnamefont{Hagen}},
  \bibinfo{author}{\bibfnamefont{D.}~\bibnamefont{Hierl}},
  \bibinfo{author}{\bibfnamefont{C.~B.} \bibnamefont{Lang}}, \bibnamefont{and}
  \bibinfo{author}{\bibfnamefont{A.}~\bibnamefont{Sch{\"a}fer}},
  \bibinfo{journal}{PoS} \textbf{\bibinfo{volume}{LAT2005}},
  \bibinfo{pages}{75} (\bibinfo{year}{2005}{\natexlab{b}}),
  \eprint{hep-lat/0509051}.

\bibitem[{\citenamefont{Burch et~al.}(2005{\natexlab{c}})\citenamefont{Burch,
  Gattringer, Glozman, Hagen, Hierl, Lang, and Sch{\"a}fer}}]{BuGaGl05c}
\bibinfo{author}{\bibfnamefont{T.}~\bibnamefont{Burch}},
  \bibinfo{author}{\bibfnamefont{C.}~\bibnamefont{Gattringer}},
  \bibinfo{author}{\bibfnamefont{L.~Y.}~\bibnamefont{Glozman}},
  \bibinfo{author}{\bibfnamefont{C.}~\bibnamefont{Hagen}},
  \bibinfo{author}{\bibfnamefont{D.}~\bibnamefont{Hierl}},
  \bibinfo{author}{\bibfnamefont{C.~B.} \bibnamefont{Lang}}, \bibnamefont{and}
  \bibinfo{author}{\bibfnamefont{A.}~\bibnamefont{Sch{\"a}fer}},
  \bibinfo{journal}{PoS} \textbf{\bibinfo{volume}{LAT2005}},
  \bibinfo{pages}{097} (\bibinfo{year}{2005}{\natexlab{c}}),
  \eprint{hep-lat/0509086}.

\bibitem[{\citenamefont{Burch et~al.}(2006{\natexlab{a}})\citenamefont{Burch,
  Gattringer, Glozman, Hagen, and Lang}}]{BuGaGl06a}
\bibinfo{author}{\bibfnamefont{T.}~\bibnamefont{Burch}},
  \bibinfo{author}{\bibfnamefont{C.}~\bibnamefont{Gattringer}},
  \bibinfo{author}{\bibfnamefont{L.~Y.} \bibnamefont{Glozman}},
  \bibinfo{author}{\bibfnamefont{C.}~\bibnamefont{Hagen}}, \bibnamefont{and}
  \bibinfo{author}{\bibfnamefont{C.~B.} \bibnamefont{Lang}},
  \bibinfo{journal}{Phys. Rev. D} \textbf{\bibinfo{volume}{73}},
  \bibinfo{pages}{017502} (\bibinfo{year}{2006}{\natexlab{a}}),
  \eprint{hep-lat/0511054}.

\bibitem[{\citenamefont{Burch et~al.}(2005{\natexlab{d}})\citenamefont{Burch,
  Gattringer, Glozman, Kleindl, Lang, , and Sch{\"a}fer}}]{BuGaGl05d}
\bibinfo{author}{\bibfnamefont{T.}~\bibnamefont{Burch}},
  \bibinfo{author}{\bibfnamefont{C.}~\bibnamefont{Gattringer}},
  \bibinfo{author}{\bibfnamefont{L.~Y.} \bibnamefont{Glozman}},
  \bibinfo{author}{\bibfnamefont{R.}~\bibnamefont{Kleindl}},
  \bibinfo{author}{\bibfnamefont{C.~B.} \bibnamefont{Lang}},
  \bibnamefont{and}
  \bibinfo{author}{\bibfnamefont{A.}~\bibnamefont{Sch{\"a}fer}},
  \bibinfo{journal}{Nucl. Phys.} \textbf{\bibinfo{volume}{A755}},
  \bibinfo{pages}{481} (\bibinfo{year}{2005}{\natexlab{d}}),
  \eprint{nucl-th/0501025}.

\bibitem[{\citenamefont{Burch et~al.}(2006{\natexlab{b}})\citenamefont{Burch,
  Gattringer, Glozman, Hagen, Lang, and Sch{\"a}fer}}]{BuGaGl06}
\bibinfo{author}{\bibfnamefont{T.}~\bibnamefont{Burch}},
  \bibinfo{author}{\bibfnamefont{C.}~\bibnamefont{Gattringer}},
  \bibinfo{author}{\bibfnamefont{L.~Y.} \bibnamefont{Glozman}},
  \bibinfo{author}{\bibfnamefont{C.}~\bibnamefont{Hagen}},
  \bibinfo{author}{\bibfnamefont{C.~B.} \bibnamefont{Lang}}, \bibnamefont{and}
  \bibinfo{author}{\bibfnamefont{A.}~\bibnamefont{Sch{\"a}fer}},
  \bibinfo{journal}{Phys. Rev. D} \textbf{\bibinfo{volume}{73}},
  \bibinfo{pages}{094505} (\bibinfo{year}{2006}{\natexlab{b}}),
  \eprint{hep-lat/0601026}.

\bibitem[{\citenamefont{Burch et~al.}(2006{\natexlab{c}})\citenamefont{Burch,
  Gattringer, Glozman, Hagen, Hierl, Lang, and Sch{\"a}fer}}]{BuGaGl06b}
\bibinfo{author}{\bibfnamefont{T.}~\bibnamefont{Burch}},
  \bibinfo{author}{\bibfnamefont{C.}~\bibnamefont{Gattringer}},
  \bibinfo{author}{\bibfnamefont{L.~Y.} \bibnamefont{Glozman}},
  \bibinfo{author}{\bibfnamefont{C.}~\bibnamefont{Hagen}},
  \bibinfo{author}{\bibfnamefont{D.}~\bibnamefont{Hierl}},
  \bibinfo{author}{\bibfnamefont{C.~B.} \bibnamefont{Lang}}, \bibnamefont{and}
  \bibinfo{author}{\bibfnamefont{A.}~\bibnamefont{Sch{\"a}fer}},
  \bibinfo{journal}{Phys. Rev. D} \textbf{\bibinfo{volume}{74}},
  \bibinfo{pages}{014504} (\bibinfo{year}{2006}{\natexlab{c}}),
  \eprint{hep-lat/0604019}.

\bibitem[{\citenamefont{Gattringer et~al.}(2008)\citenamefont{Gattringer,
  Glozman, Lang, Mohler, and Prelovsek}}]{GaGlLa08}
\bibinfo{author}{\bibfnamefont{C.}~\bibnamefont{Gattringer}},
  \bibinfo{author}{\bibfnamefont{L.~Y.} \bibnamefont{Glozman}},
  \bibinfo{author}{\bibfnamefont{C.~B.} \bibnamefont{Lang}},
  \bibinfo{author}{\bibfnamefont{D.}~\bibnamefont{Mohler}}, \bibnamefont{and}
  \bibinfo{author}{\bibfnamefont{S.}~\bibnamefont{Prelovsek}},
  \bibinfo{journal}{Phys. Rev. D} \textbf{\bibinfo{volume}{78}},
  \bibinfo{pages}{034501} (\bibinfo{year}{2008}), \eprint{arXiv:0802.2020
  [hep-lat]}.

\bibitem[{\citenamefont{Morningstar}(2008)}]{Mo08}
\bibinfo{author}{\bibfnamefont{C.}~\bibnamefont{Morningstar}}
  (\bibinfo{year}{2008}), \eprint{arXiv:0810.4448 [hep-lat]}.

\bibitem[{\citenamefont{Blossier et~al.}(2008)\citenamefont{Blossier, von
  Hippel, Mendes, Sommer, and DellaMorte}}]{BlHiMe08}
\bibinfo{author}{\bibfnamefont{B.}~\bibnamefont{Blossier}},
  \bibinfo{author}{\bibfnamefont{G.}~\bibnamefont{von Hippel}},
  \bibinfo{author}{\bibfnamefont{T.}~\bibnamefont{Mendes}},
  \bibinfo{author}{\bibfnamefont{R.}~\bibnamefont{Sommer}}, \bibnamefont{and}
  \bibinfo{author}{\bibfnamefont{M.}~\bibnamefont{DellaMorte}},
  \bibinfo{journal}{PoS} \textbf{\bibinfo{volume}{LATTICE2008}},
  \bibinfo{pages}{135} (\bibinfo{year}{2008}),
\eprint{arXiv:0808.1017 [hep-lat]}.

\bibitem[{\citenamefont{G{\"u}sken et~al.}(1989)}]{Gu89}
\bibinfo{author}{\bibfnamefont{S.}~\bibnamefont{G{\"u}sken}}
  \bibnamefont{et~al.}, \bibinfo{journal}{Phys. Lett. B}
  \textbf{\bibinfo{volume}{227}}, \bibinfo{pages}{266} (\bibinfo{year}{1989}).

\bibitem[{\citenamefont{Best et~al.}(1997)}]{Be97}
\bibinfo{author}{\bibfnamefont{C.}~\bibnamefont{Best}} \bibnamefont{et~al.},
  \bibinfo{journal}{Phys. Rev. D} \textbf{\bibinfo{volume}{56}},
  \bibinfo{pages}{2743} (\bibinfo{year}{1997}), \eprint{hep-lat/9703014}.

\bibitem[{\citenamefont{Lichtl}(2007)}]{Li07a}
\bibinfo{author}{\bibfnamefont{A.~C.} \bibnamefont{Lichtl}},
  \bibinfo{journal}{PoS} \textbf{\bibinfo{volume}{LATTICE2007}},
  \bibinfo{pages}{118} (\bibinfo{year}{2007}),
\eprint{arXiv:0711.4072 [hep-lat]}.

\bibitem[{\citenamefont{Danzer and Gattringer}(2007)}]{DaGa07}
\bibinfo{author}{\bibfnamefont{J.}~\bibnamefont{Danzer}} \bibnamefont{and}
  \bibinfo{author}{\bibfnamefont{C.}~\bibnamefont{Gattringer}},
  \bibinfo{journal}{PoS} \textbf{\bibinfo{volume}{LAT2007}},
  \bibinfo{pages}{092} (\bibinfo{year}{2007}), \eprint{arXiv:0710.1711
  [hep-lat]}.

\bibitem[{\citenamefont{Gattringer
  et~al.}(2001{\natexlab{b}})\citenamefont{Gattringer, G{\"o}ckeler, Rakow,
  Schaefer, and Sch{\"a}fer}}]{GaGoRa01b}
\bibinfo{author}{\bibfnamefont{C.}~\bibnamefont{Gattringer}},
  \bibinfo{author}{\bibfnamefont{M.}~\bibnamefont{G{\"o}ckeler}},
  \bibinfo{author}{\bibfnamefont{P.~E.~L.} \bibnamefont{Rakow}},
  \bibinfo{author}{\bibfnamefont{S.}~\bibnamefont{Schaefer}}, \bibnamefont{and}
  \bibinfo{author}{\bibfnamefont{A.}~\bibnamefont{Sch{\"a}fer}},
  \bibinfo{journal}{Nucl. Phys.} \textbf{\bibinfo{volume}{B618}},
  \bibinfo{pages}{205} (\bibinfo{year}{2001}{\natexlab{b}}),
  \eprint{hep-lat/0105023}.

\bibitem[{\citenamefont{Alford et~al.}(1995)\citenamefont{Alford, Dimm, Lepage,
  Hockney, and Mackenzie}}]{AlDiLe95}
\bibinfo{author}{\bibfnamefont{M.}~\bibnamefont{Alford}},
  \bibinfo{author}{\bibfnamefont{W.}~\bibnamefont{Dimm}},
  \bibinfo{author}{\bibfnamefont{G.~P.} \bibnamefont{Lepage}},
  \bibinfo{author}{\bibfnamefont{G.}~\bibnamefont{Hockney}}, \bibnamefont{and}
  \bibinfo{author}{\bibfnamefont{P.~B.} \bibnamefont{Mackenzie}},
  \bibinfo{journal}{Phys. Lett. B} \textbf{\bibinfo{volume}{361}},
  \bibinfo{pages}{87} (\bibinfo{year}{1995}),
\eprint{hep-lat/9507010}.

\end{thebibliography}

\end{document}